%% file: paper.tex
\documentclass[nohyperref]{article}
\usepackage[a4paper, total={6in, 8in}]{geometry}
\usepackage{microtype}
\usepackage{graphicx}
\usepackage{wrapfig}
\usepackage{subcaption}
\usepackage{booktabs}
\usepackage{multirow}
\usepackage{authblk}
\usepackage[ruled]{algorithm2e}

\usepackage{listings}
\usepackage{subcaption}
\usepackage{biblatex}
\usepackage{hyperref}

\usepackage{amsmath}
\usepackage{amssymb}
\usepackage{mathtools}
\usepackage{amsthm}

\usepackage{makecell}

\usepackage[capitalize,noabbrev]{cleveref}

\usepackage[skip=10pt plus1pt, indent=0pt]{parskip}

\theoremstyle{plain}

\newtheorem{example}{Example}
\theoremstyle{definition}

\theoremstyle{remark}

\newcommand{\imgZERONE}{%
  \begingroup\normalfont
  \includegraphics[height=1.5\fontcharht\font`\B]{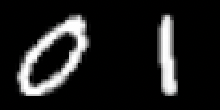}%
  \endgroup
}

\usepackage[textsize=tiny]{todonotes}

\title{VAEL: Bridging Variational Autoencoders and Probabilistic Logic Programming.}

\author[1]{Eleonora Misino\thanks{eleonora.misino2@unibo.it}}
\author[2]{Giuseppe Marra\thanks{giuseppe.marra@kuleuven.be}}
\author[2]{Emanuele Sansone\thanks{emanuele.sansone@kuleuven.be}}
\affil[1]{DISI, University of Bologna}
\affil[2]{Department of Computer Science, KU Leuven}
\date{}                   
\begin{document}

\maketitle

\begin{abstract}
We present VAEL, a neuro-symbolic generative model integrating variational autoencoders (VAE) with the reasoning capabilities of probabilistic logic (L) programming.  Besides standard latent subsymbolic variables, our model exploits a probabilistic logic program to define a further structured representation, which is used for logical reasoning. The entire process is end-to-end differentiable. Once trained, VAEL can solve new unseen generation tasks by (i) leveraging the previously acquired knowledge encoded in the neural component and (ii) exploiting new logical programs on the structured latent space. Our experiments provide support on the benefits of this neuro-symbolic integration both in terms of task generalization and data efficiency. To the best of our knowledge, this work is the first to propose a general-purpose end-to-end framework integrating probabilistic logic programming into a deep generative model.
\end{abstract}

\section{Introduction}
Neuro-symbolic learning has gained tremendous attention in the last few years \cite{besold2017neural,deraedt2020statistical,kautz2020third,bengio2021aidebate} as such integration has the potential of leading to a new era of intelligent solutions, enabling the integration of deep learning and reasoning strategies (e.g. logic-based or expert systems). Indeed, these two worlds have different strengths that complement each other \cite{kahneman2011thinking}. For example, deep learning systems, i.e. System 1, excel at dealing with noisy and ambiguous high dimensional raw data, whereas reasoning systems, i.e. System 2, leverage relations between symbols to reason and to generalize from a small amount of training data. While a lot of effort has been devoted to devising neuro-symbolic methods in the discriminative setting \cite{manhaeve2018deepproblog,yi2018neural,minervini2020learning}, less attention has been paid to the generative counterpart. 
A good neuro-symbolic framework should be able to leverage a small amount of training data, acquire the knowledge by learning a symbolic representation and generate data based on new forms of high-level reasoning. For example, let us consider a task where a single image of multiple handwritten numbers is labeled with their sum. Common generative approaches, like VAE-based models, have a strong connection between the latent representation and the label of the training task~\cite{kingma2014semisupervised,joy2021capturing}.
Consequently, when considering new generation tasks that go beyond the simple addition, they have to be retrained on new data.

In this paper, we tackle the problem by providing a true neuro-symbolic solution, named VAEL. 
In VAEL the latent representation is not directly linked to the label of the task, but to a set of newly introduced symbols, i.e. logical expressions. Starting from these expressions, we use a \textit{probabilistic logic program} to deduce the label. Importantly, the neural component only needs to learn a mapping from the raw data to this new symbolic representation. In this way, the model only weakly depends on the training data and can generalize to new generation tasks involving the same set of symbols. Moreover, the reasoning component offers a strong inductive bias, which enables a more data efficient learning. 

The paper is structured as follows. In Section \ref{sec:preliminaries}, we provide a brief introduction to probabilistic logic programming and to generative models conditioned on labels. In Section \ref{sec:model}, we present the VAEL model together with its inference and learning strategies. Section \ref{sec:results} shows our experiments, while Section \ref{sec:related} places our model in the wider scenario of multiple related works. Finally, in Section \ref{sec:conclusions}, we draw some conclusions and discuss future directions.

\section{Preliminaries}

\input{src/background}

\section{The VAEL Model}

\input{src/model.tex}

\section{Experiments}
\input{src/results}

\section{Related Work}
\input{src/related}

\section{Conclusions and Future Works}
\label{sec:conclusions}
In this paper, we presented VAEL, a neuro-symbolic generative model that integrates VAE with Probabilistic Logic Programming. The symbolic component allows to decouple the internal latent representation from the task at hand, thus allowing an unprecedented generalization power. We showcased the potential of VAEL in two image generation benchmarks, where VAEL shows state-of-the-art generation performance, also in regimes of data scarcity and in generalization to several prediction tasks.

In the future, we plan to improve VAEL by investigating alternative and more scalable semantics for probabilistic programs (e.g. stochastic logic program \cite{winters2021deepstochlog}). Moreover, we plan to apply VAEL to other settings, like structured object generation \cite{liello2020efficient}, to showcase the flexibility and expressivity provided by the integration with a probabilistic logic program.

\section*{Acknowledgements}
Giuseppe Marra is funded by the Research Foundation-Flanders (FWO-Vlaanderen, GA No 1239422N). Emanuele Sansone is funded by TAILOR, a project funded by EU Horizon 2020 research and innovation programme under GA No 952215. The authors would like to thank Luc De Raedt for supporting this project as an Erasmus Master Thesis, and Federico Ruggeri for his support in the experimental phase.

\printbibliography

\newpage

\input{src/appendices.tex}

\end{document}

%% file: src/background.tex
\label{sec:preliminaries}

\subsection{Probabilistic Logic Programming}
\label{sec:problog}

A \textit{logic program} is a set of \textit{definite clauses}, i.e. expressions of the form  $ h \leftarrow b_1 \wedge ... \wedge b_n$, where
$h$ is the \textit{head literal} or conclusion, while  the $b_i$ are \textit{body literals} or conditions. Definite clauses can be seen as computational rules: IF all the body literals are true THEN the head literal is true. Definite clauses with no conditions ($n=0$) are \textit{facts}. In first-order logic programs, literals take the form  $a(t_1, ... , t_m)$, with $a$ a predicate of arity $m$ and $t_i$ are the terms, that is constants, variables or functors (i.e. functions of other terms). Grounding is the process of substituting all the variable in an atom or a clause with constants.

ProbLog~\cite{deraedt2007problog} lifts logic programs to \textit{probabilistic logic programs} through the introduction of probabilistic facts. Whereas a fact in a logic program is deterministically true, a probabilistic fact is of the form $p_i::f_i$
where $f_i$ is a logical fact and $p_i$ is a probability. 
In ProbLog, each ground instance  of a probabilistic fact $f_i$ corresponds to an \textit{independent Boolean random variable} that is true with probability $p_i$ and false with probability $1-p_i$. Mutually exclusive facts can be defined through \textit{annotated disjunctions} $p_0::f_0;~...~;p_n::f_n.$ with $\sum_i p_i = 1$.
Let us denote with $\mathcal{F}$ the set of all ground instances of probabilistic facts and with $p$ their corresponding probabilities. Every subset $F \subseteq \mathcal{F}$ defines a \textit{possible world} $w_F$ obtained by adding to $F$ all the atoms that can be derived from $F$ using the logic program. The probability $P(w_F;p)$ of such a possible world $w_F$ is given by the product of the probabilities of the truth values of the probabilistic facts; i.e:
\begin{equation}
P(w_F;p) = \prod_{f_i \in F}p_i \prod_{f_i \in \mathcal{F} \setminus F}\left(1 - p_i\right)
\label{eq:p_worlds}
\end{equation}
Two inference tasks on these probabilities are of interest for this paper.

\textbf{Success}: The probability of a query atom $y$, or formula, also called \textit{success probability of $y$}, is the sum of the probabilities of all worlds where $y$ is \textit{True}, i.e.,
\begin{equation}
P(y;p) = \sum_{F \subseteq \mathcal{F} : w_F \models y}  P(w_F;p)
\label{eq:p_query}
\end{equation}
\textbf{Sample with evidence}: Given a set of atoms or formulas $E$, the \textit{evidence}, the probability of a world given evidence is:
\begin{equation}
P(w_F | E;p) = \frac{1}{Z}\begin{cases}
                        P(w_F;p) & \text{if~} w_F \models E \\
                        0 & otherwise
                    \end{cases}
\label{eq:p_cond}
\end{equation}
where $Z$ is a normalization constant. Sampling from this distribution provides only worlds that are coherent with the given evidence.

\begin{example}[Addition of two digits]
\label{ex:digits_add} Let us consider a setting where images contains two digits that can only be $0$ or $1$. 
Consider the following two logical predicates: $\mathtt{digit(img,I,Y)}$ states that a given image $\mathtt{img}$ has  a certain digit $\mathtt{Y}$ in position $\mathtt{I}$, while $\mathtt{add(img,z)}$ states that the digits in $\mathtt{img}$ sum to a certain value $z$.

We can encode the digit addition task in the following program $T$:

\begin{lstlisting}[basicstyle = { \ttfamily \footnotesize} ]
p1::digit(img,1,0); p2::digit(img,1,1). 
p3::digit(img,2,0); p4::digit(img,2,1). 

add(img,Z) :- digit(img,1,Y1),
              digit(img,2,Y2), 
              Z is Y1 + Y2.
\end{lstlisting}
In this program $T$, the set of ground facts $\mathcal{F}$ is  $$\{\mathtt{digit(img,1,0), digit(img,1,1),digit(img,2,0),digit(img,2,1)}\}.$$ The set of probabilities $p$ is $p = [p_1, p_2, p_3, p_4]$. The ProbLog program $T$ defines a probability distribution over the possible worlds and it is parameterized by $p$, i.e. $P(\omega_F;p)$. 
Then, we can ask ProbLog to  compute the  success probability of a query using Equation \ref{eq:p_query}, e.g. $P(\mathtt{add(img,1)})$; or sample a possible world coherent with some evidence $\mathtt{add(img,2)}$ using Equation \ref{eq:p_cond}, e.g. $w_F = \{\mathtt{digit(img,1,1)}, \mathtt{digit(img,2,1)}\}$.

\end{example}

\subsection{Generation Conditioned on Labels}

\begin{figure}[h]
     \centering
     \begin{subfigure}[b]{0.33\linewidth}
         \centering
         \includegraphics[width=0.7\textwidth]{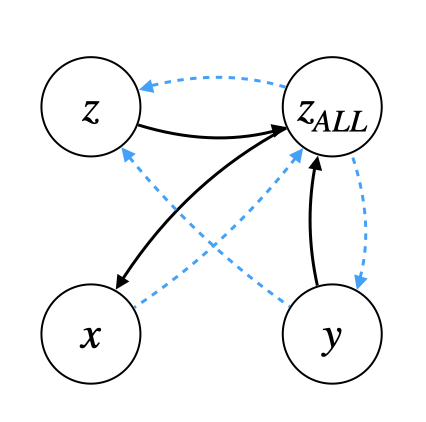}
         \caption{M1+M2}
     \end{subfigure}%
     \begin{subfigure}[b]{0.33\linewidth}
         \centering
         \includegraphics[width=0.7\textwidth]{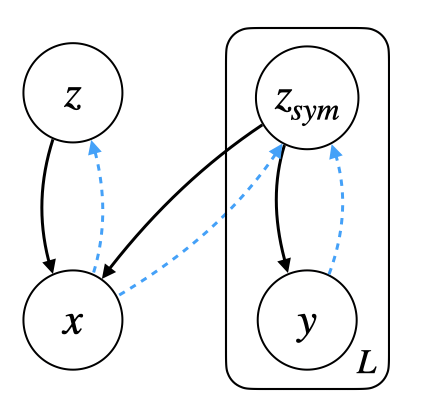}
         \caption{CCVAE}
     \end{subfigure}%
     \begin{subfigure}[b]{0.33\linewidth}
         \centering
         \includegraphics[width=0.7\textwidth]{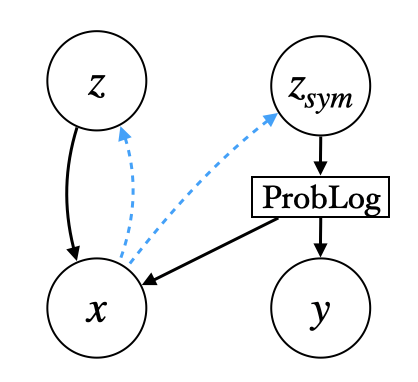}
         \caption{VAEL (ours)}
         \label{fig:pgm_VAEL}
     \end{subfigure}%
  \caption{Visual comparison for the probabilistic graphical models of~\cite{kingma2014semisupervised} (M1+M2), of~\cite{joy2021capturing} (CCVAE) and ours (VAEL). Black arrows refer to the generative model, whereas blue dashed arrows correspond to the inference counterpart.}

     \label{fig:graphical_models}

\end{figure}

In this paper, we are interested in generative tasks where we consider both an image $x$ and a label $y$.
The integration of supervision into a generative latent variable model has been largely investigated in the past. For example, the work of~\cite{kingma2014semisupervised} proposes an integrated framework between two generative models, called M1 and M2 (cf. Figure~\ref{fig:graphical_models}). Model M1 learns a latent representation for input $x$, i.e. $z_{ALL}$, which is further decomposed by model M2 into a symbolic and a subsymbolic vector $y$ and $z$, respectively. In this formulation, the generative process of the image is tightly dependent on the label, and therefore on the training task. More recently, another approach, called 

CCVAE \cite{joy2021capturing}, proposes to learn a representation consisting of two independent latent vectors, i.e. $z$ and $z_{sym}$, and forces the elements of $z_{sym}$ to have a one-to-one correspondence with the $L$ elements of $y$, thus capturing the rich information of the label vector $y$ (cf. Figure~\ref{fig:graphical_models}).

However, both the approaches are limited in terms of generation ability as their latent representation encodes information about the training task. This could be problematic when the label $y$ is only weakly linked to the true symbolic structure of the image. For example, let us consider the addition task in Example \ref{ex:digits_add}, where a single image of multiple handwritten numbers is  labeled with their sum, e.g. $x = $~\imgZERONE ~ and $y = 1$.  In a generative task where we are interested in creating new images, using only the information of the label $y$ is not as expressive as directly using the values of the single digits. Moreover, suppose that we want to generate images where the two digits are related by other operations (e.g. subtraction, multiplication, etc). While we still want to generate an image representing a pair of digits, none of the models mentioned before would be able to do it without being retrained on a relabelled dataset. How can we overcome such limitations?

%% file: src/model.tex
\label{sec:model}

\begin{figure*}[htp]
\centering
  \includegraphics[width=0.7\textwidth]{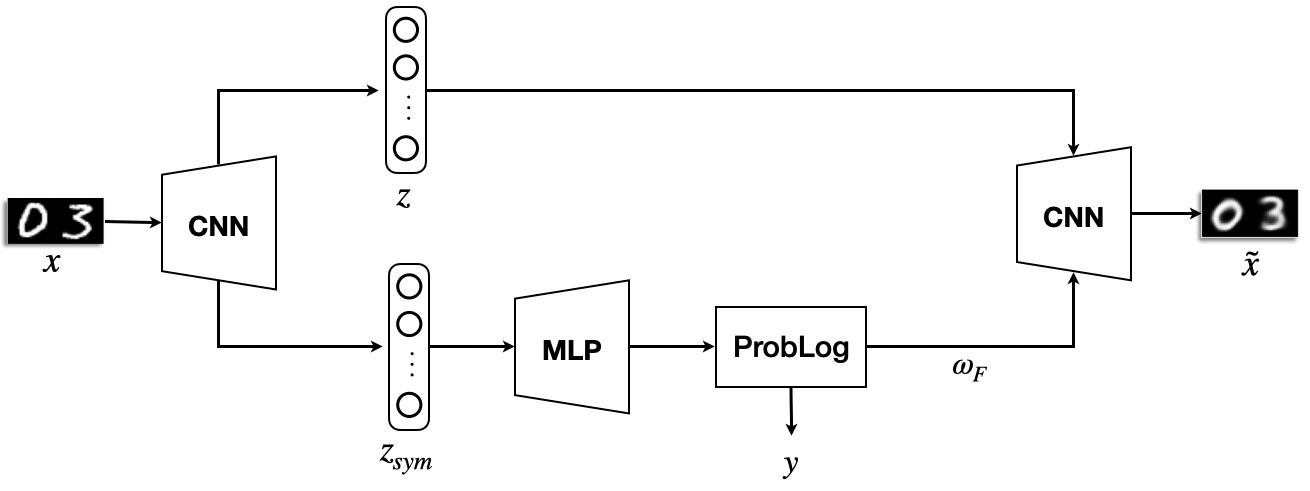}
  \caption{The VAEL model is composed of three components. First, the encoder (\textbf{left}) computes an approximated posterior of the latent variables $\mathbf{z}$ from the image $x$. The latent variables are split into two components: a subsymbolic $z$ and a symbolic $z_{sym}$.  Second, $z_{sym}$ is used to parameterize a ProbLog program (\textbf{center}). A MLP is used to map the real variables $z_{sym}$ into the probabilities of the facts in the program. Then, the program is used to compute the label $y$ and a possible world. Finally, a decoder (\textbf{right}) takes both the latent vector $z$ and the possible world from ProbLog to reconstruct the image $\tilde x$.}
\label{fig:vael_architecture} 
\end{figure*}

Here, we propose a probabilistic graphical model which enables to unify VAEs with Probabilistic Logic Programming. The graphical model of VAEL (Figure \ref{fig:graphical_models}) consists of four core variables. $x \in \mathbb{R}^{H \times W \times C} $ represents the image we want to generate, while $y \in \{0,1\}^K$ represents a label, i.e. a symbolic information characterizing the image. The latent variable is split into a symbolic component $z_{sym} \in \mathbb{R}^N$ and a subsymbolic component $z \in \mathbb{R}^M$.
 Conversely to other VAE frameworks, VAEL does not rely on a one-to-one mapping between $y$ and $z_{sym}$, rather it exploits a probabilistic logic program to link them. Indeed, the probabilistic facts $\mathcal{F}$ are used by the ProbLog program $T$ to compute the actual labels $y$ and they can encode a more meaningful symbolic representation of the image than $y$.

\textbf{Generative model.} 

The generative distribution of VAEL (Figure \ref{fig:graphical_models}) is factorized in the following way:
\begin{equation}
    p_{\theta}(x,y,\mathbf{z}) = p(x|\mathbf{z})p(y|z_{sym})p(\mathbf{z})
    \label{eq:generative_network}
\end{equation}
where $\mathbf{z} = [z_{sym}, z]$ and $\theta$ are the parameters of the generative model. $p(\mathbf{z})$ is a standard Gaussian distribution, while $p(y|z_{sym})$ is the success distribution of the label of the ProbLog program $T$ (Eq. \ref{eq:p_query}). $p(x|\mathbf{z})$ is a Laplace distribution with mean value $\mu$ and identity covariance, i.e. 
$Laplace(x; \mu, I)$. Here, $\mu$ is a neural network decoder whose inputs are $z$ and $\omega_F$. $\omega_F$ is sampled from $P(\omega_F; MLP(z_{sym}))$ (Eq. \ref{eq:p_worlds}).

\textbf{Inference model.} We amortise inference by using an approximate posterior distribution $q_\phi(\mathbf{z}|x,y)$ with parameters $\phi$. Furthermore, we assume that $\mathbf{z}$ and $y$ are conditionally independent given $x$, thus obtaining $q_\phi(\mathbf{z}|x,y) = q_{\phi}(\mathbf{z}|x)$\footnote{We use a Gaussian distribution with a mean parameterized by the encoder network and identity covariance}. This allows us to decouple the latent representation from the training task.
Conversely, the other VAE frameworks do not exploit this assumption and have a latent representation that is dependent on the training task.

The overall VAEL model (including the inference and the generative components) is shown in Figure~\ref{fig:vael_architecture}.

\textbf{Objective Function.} The objective function of VAEL computes an evidence lower bound (ELBO) on the log likelihood of pair $(x,y)$, namely:
\begin{equation}
\mathcal{L}(\theta,\phi) = \mathcal{L}_{REC}( \theta,\phi) + \mathcal{L}_{Q}( \theta,\phi) -
\mathcal{D_{KL}}[q_{\phi}(\mathbf{z}|x)||p(\mathbf{z})] ] \label{eq:loss}
\end{equation}
where
$$ 
\mathcal{L}_{REC}( \theta,\phi) = \mathbb{E}_{\mathbf{z}\sim q_{\phi}(\mathbf{z}|x)}[\log(p(x|\mathbf{z})], \quad
\mathcal{L}_{Q}(\theta,\phi) = 
     \mathbb{E}_{z_{sym}\sim q_{\phi}(z_{sym}|x))}[
       \log(p(y|z_{sym}))]].
$$

Note that we omit the dependence on $\omega_F$ in the objective, thanks to an equivalence described in the extended derivation (see Appendix \ref{app:elbo_derivation}).

The objective is used to train VAEL in an end-to-end differentiable manner, thanks to the Reparametrization Trick~\cite{kingma2013autoencoding} at the level of the encoder  $q_{\phi}(\mathbf{z}|x)$ and the differentiability of the ProbLog inference, which is used to compute the success probability of a query and sample a world.

In Appendix \ref{app:learning} we report VAEL training algorithm (Algorithm \ref{training_alg}) along with further details on the training procedure.

\subsection{Downstream Applications}
\textbf{Label Classification.} Given $x$ we use the encoder to compute $z_{sym}$ and by using the MLP we compute the probabilities $p = MLP(z_{sym})$. Then, we can predict labels by computing the probability distribution over the labels $P(y; p)$, as defined in Eq. \ref{eq:p_query}, and sampling $y \sim P(y; p)$. This process subsumes the DeepProbLog framework \cite{manhaeve2018deepproblog}.

\textbf{Image Generation.} We generate images by sampling $\mathbf{z} = [z_{sym},z]$ from the prior distribution $\mathcal{N}(0,1)$ and a possible world  $\omega_{F}$ from $P(\omega_F;p)$. The distribution over the possible worlds $P(\omega_F;p)$ is computed by relying on ProbLog inference starting from the facts probabilities $p = MLP(z_{sym})$.

\textbf{Conditional Image Generation.}
As described in Section \ref{sec:problog}, ProbLog inference allows us also to \textit{sample with evidence}. Thus, once sampled $\mathbf{z}$ from the prior, we can (i) compute $p = MLP(\hat{z}_{sym})$, then  (ii) compute the conditional probability $P(\omega_F \mid E; p)$, (iii) sampling $\omega_F \sim P(\omega_F \mid E; p)$ and (iv) generate an image consistent with the evidence $E$.

\textbf{Task Generalization}
As we have seen, VAEL factorizes the generation task into two steps: (i) generation of the world $\omega_F$ (e.g. the digits labels); (ii) generation of the image given the world. Whereas the second step requires to be parameterized by a black-box model (e.g. a convolutional neural network), the generation of a possible world $\omega_F$ is handled by a symbolic generative process encoded in the ProbLog program $T$. 
Thus, once trained VAEL on a specific symbolic task (e.g. the addition of two digits), we can generalize to any novel task that involves reasoning with the same set of probabilistic facts by simply changing the ProbLog program accordingly (e.g. we can generalize to the multiplication of two integers). To the best of our knowledge, such a level of task generalization cannot be achieved by any other VAE frameworks.

%% file: src/results.tex
\label{sec:results}
In this Section, we validate our approach on the four downstream applications by creating two different datasets. 

\textbf{2digit MNIST dataset.} We create a dataset of $64,400$ images of two digits taken from the MNIST dataset \cite{lecun1998mnist}. We use 65\%, 20\%, 15\% splits for the train, validation and test sets, respectively. Each image in the dataset has dimension $28 \times 56$ and is labelled with the sum of the two digits. The dataset contains a number of images similar to the standard MNIST dataset. However, it is combinatorial in nature, making any task defined on it harder than its single-digit counterpart. 

\begin{wrapfigure}[14]{r}{0.3\textwidth}
\centering
    \includegraphics[width=1.\linewidth]{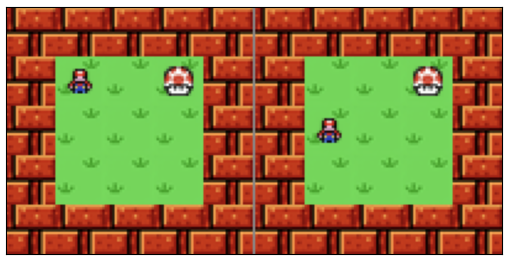}
      \caption{Example of \textit{Mario} dataset image. The $3\times3$ grid world (green area) is surrounded by a frame (bricks).}
\label{fig:mario_example}
\end{wrapfigure}
\textbf{Mario dataset.} We create a dataset containing $6,720$ images of two consequent states of a $3\times3$ grid world where an agent can move by one single step (diagonals excluded). Each image has dimension $100 \times 200$ and is labelled with the move performed by the agent. For example, the image in Figure \ref{fig:mario_example} has label \texttt{down}. We use 70\%, 20\%, 10\% splits for the train, validation and test sets, respectively. 

In order to evaluate our approach, we rely on a \textit{reconstruction loss} ($m_{REC}$) in terms of data log-likelihood and two accuracies, \textit{predictive}  ($m_{CLASS}$) and \textit{generative}  ($m_{GEN}$). Regarding the \textit{predictive accuracy}, we measure the predictive ability of the model as the classification accuracy on the true labels (the \textit{addition} of the two digits for \textit{2digit MNIST} dataset, and the \textit{move} for \textit{Mario} dataset). It is worth mentioning that, for \textit{2digit MNIST} dataset, such accuracy cannot be directly compared with standard values for the single-digit MNIST, as the input space is different: the correct classification of an image requires both the digits to be correctly classified. The \textit{generative accuracy} is assessed by using an independent classifier for each dataset. For \textit{2digit MNIST} dataset, the classifier is trained to classify single digit value; while for the \textit{Mario} dataset, the classifier learns to identify the agent's position in a single state. The evaluation process for the generative ability can be summarized as: (i) jointly generate the image and the label $\tilde{y}$; (ii) split the image into two sub-images and (iii) classify them independently; (iv) finally, for \textit{2digit MNIST} dataset, we sum together the outputs of the classifier and we compare the resulting addition with the generated label $\tilde{y}$; while for \textit{Mario Dataset}, we verify whether the classified agent's positions are consistent with the generated label $\tilde{y}$.

In the following tasks, we compare VAEL against CCVAE \cite{joy2021capturing} when possible. The source code and the datasets are available at \url{https://github.com/EleMisi/VAEL} under MIT license. Further implementation details can be found in Appendix \ref{app:implem}.

\textbf{Label Classification.} 
In this task, we want to predict the correct label given the input image, as measured by the predictive accuracy $m_{CLASS}$. 
Both VAEL and CCVAE use an encoder to map the input image to a latent vector $z_{sym}$. VAEL uses ProbLog inference to predict the label $y$. In contrast, CCVAE relies on the distribution $p(y | z_{sym})$, which is parameterized by a neural network.
As shown in Table \ref{tab:base_task}, CCVAE and VAEL achieve comparable predictive accuracy in \textit{Mario} dataset.  However, VAEL generalizes better than CCVAE in \textit{2digit MNIST} dataset. The reason behind this performance gap is due to the fact that the \textit{addition} task is combinatorial in nature and CCVAE would require a larger number of training samples in order to solve it. We further investigate this aspect in the \textit{Data efficiency} experiment.

\begin{figure}[t]
     \begin{subfigure}[b]{0.4\textwidth}
         \centering
         \includegraphics[width=0.9\textwidth]{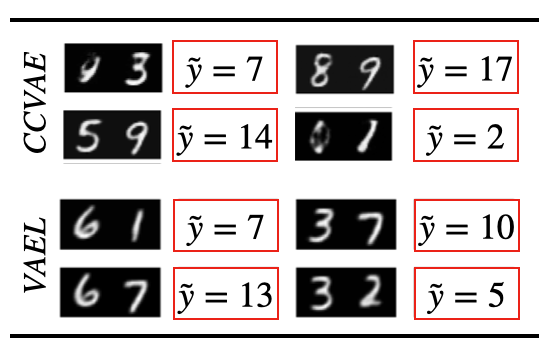}
         \caption{}
         \label{fig:gen_digits}
     \end{subfigure}
     \begin{subfigure}[b]{0.6\textwidth}
         \centering
         \includegraphics[width=0.805\textwidth]{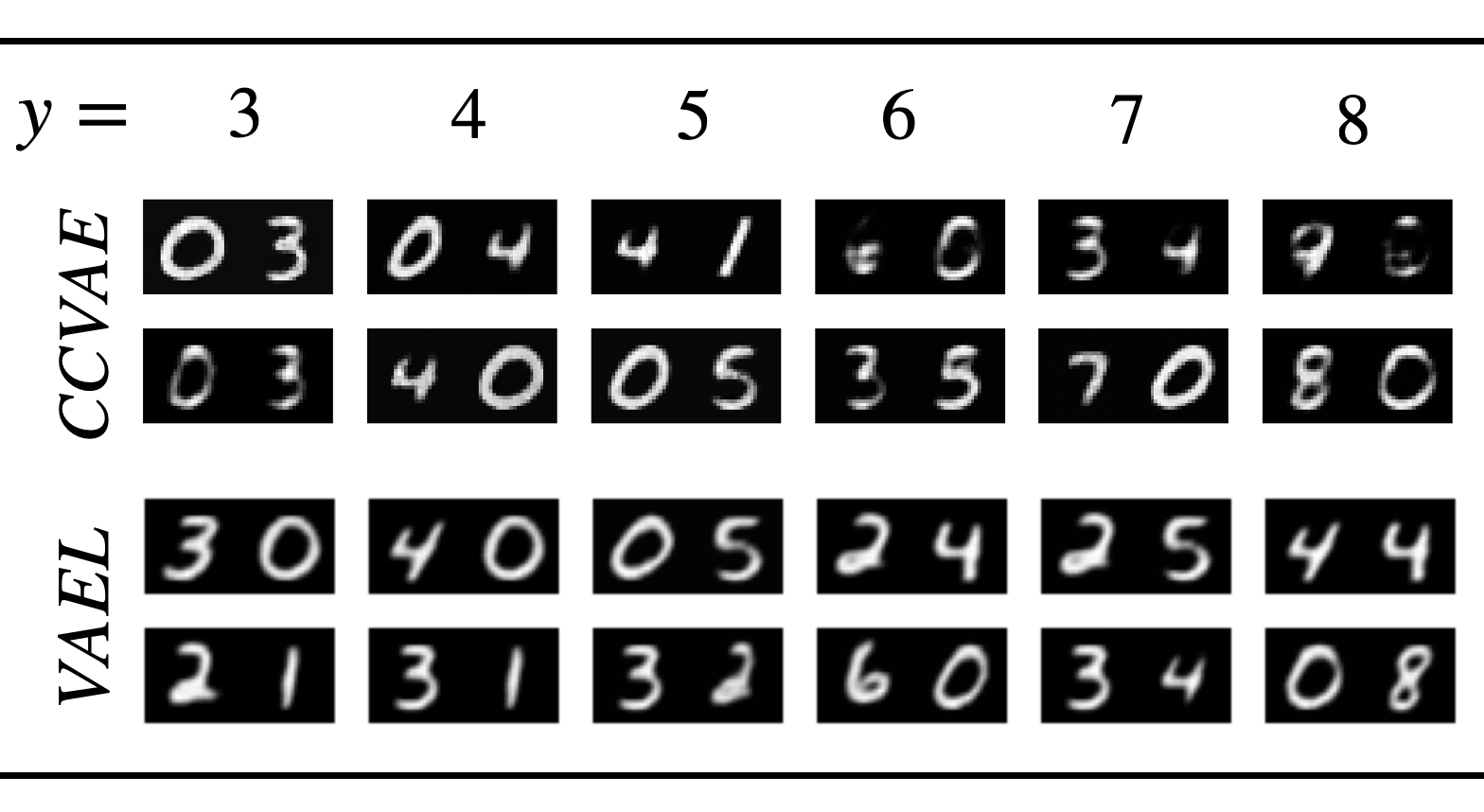}
         \caption{}
         \label{fig:cond_gen_digits}
     \end{subfigure}%
  \caption{Examples of generation (a) and conditional generation (b) for VAEL and CCVAE on \textit{2digit MNIST} dataset. In (b) in each column the generation is conditioned on a different label $y$.}

     \label{fig:digits}

\end{figure}
\begin{figure}[t]
     \centering
     \begin{subfigure}[b]{0.4\textwidth}
         \centering
         \includegraphics[width=0.75
         \textwidth]{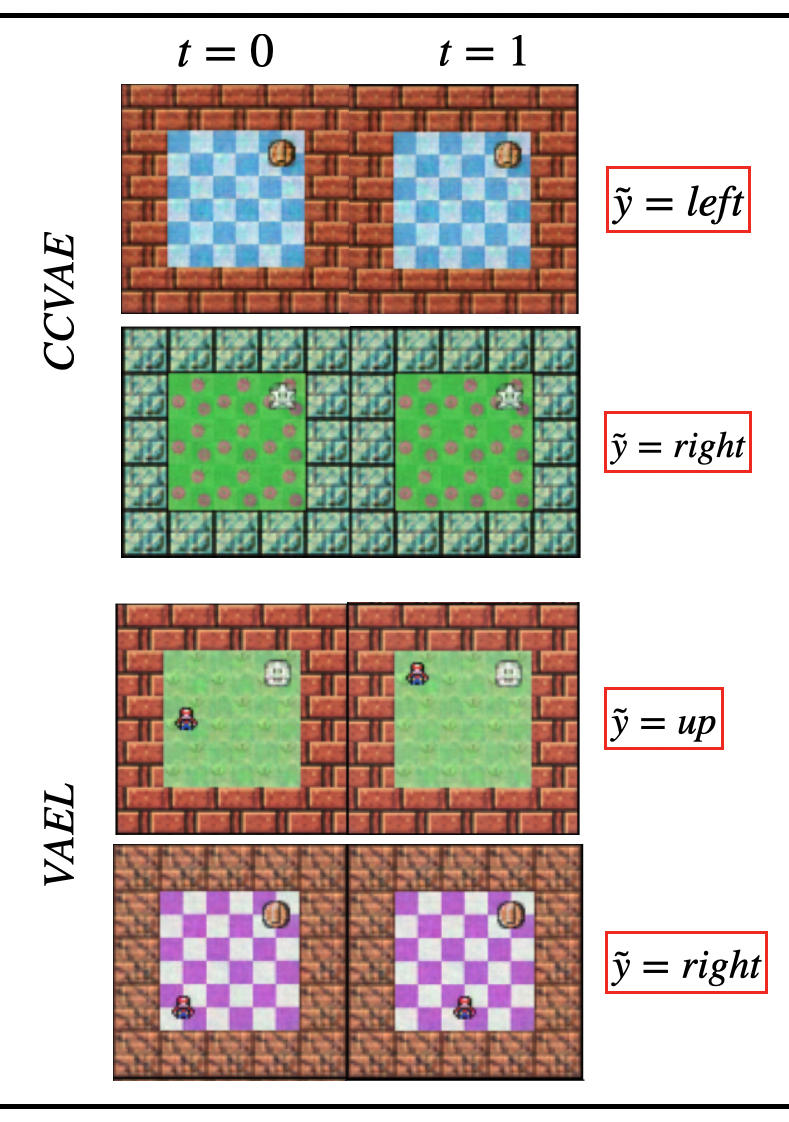}
         \caption{}
         \label{fig:gen_mario}
     \end{subfigure}%
     \begin{subfigure}[b]{0.6\textwidth}
         \centering
         \includegraphics[width=0.815\textwidth]{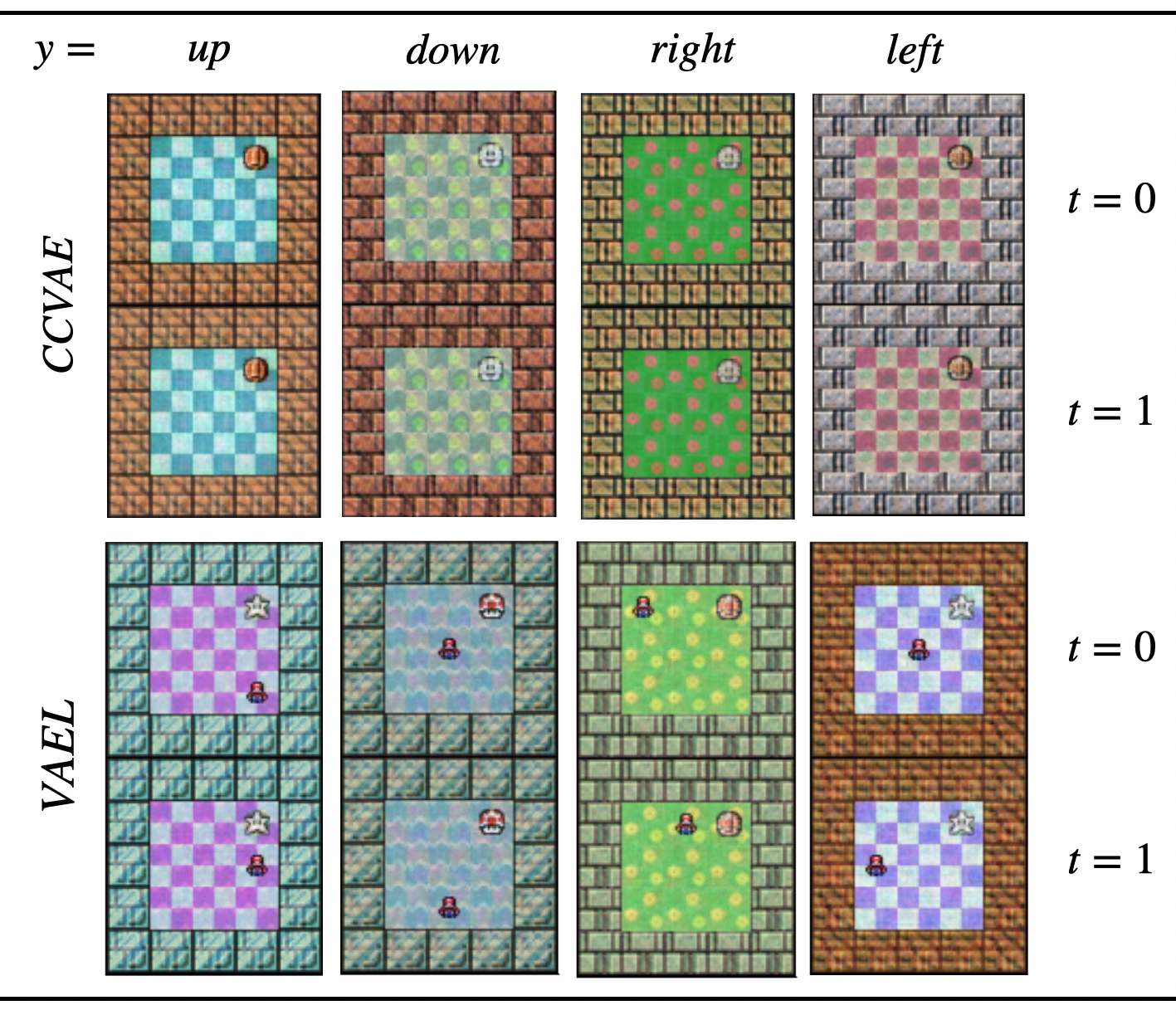}
         \caption{}
         \label{fig:cond_gen_mario}
     \end{subfigure}%
  \caption{Examples of generation (a) and conditional generation (b) for VAEL and CCVAE on \textit{Mario} dataset. In (b) in each column the generation is conditioned on a different label $y$.}

     \label{fig:mario}

\end{figure}

\textbf{Image Generation.} We want to test the performance when generating both the image and the label. VAEL generates both the image and the label $\tilde{y}$ starting from the sampled latent vector $\mathbf{z} \sim \mathcal{N}(0,1)$. Conversely, CCVAE starts by sampling the label $\tilde{y}$ from its prior, then proceeds by sampling the latent vector from $p(\mathbf{z}|y=\tilde{y})$, and finally generates the new image. Figure \ref{fig:gen_digits} shows some random samples for both models for \textit{2digit MNIST} dataset. The pairs drawn by VAEL are well defined, while CCVAE generates more ambiguous digits (e.g., the $1$ resembles a $0$, the $4$ may be interpreted as a $9$, and so on). This ambiguity makes it harder for the classifier network to distinguish among the digits during the evaluation process, as confirmed by the quantitative results in Table  \ref{tab:base_task}, where VAEL outperforms CCVAE in terms of generative ability. Regarding \textit{Mario} dataset (Figure \ref{fig:gen_mario}), VAEL is able to generate data-like images, where the background is preserved from one state to the subsequent one (additional results can be found in Appendix \ref{app:add_results}). Conversely, CCVAE fails the generation task: although it correctly generates the background, it is not able to draw the agent. This is also supported by the disparity in the reconstructive ability, as reported in Table \ref{tab:base_task}. In \textit{Mario} dataset, this is due to a systematic error in which CCVAE focuses only on reconstructing the background, thus discarding the small portion of the image containing the agent, as shown in Figures \ref{fig:gen_mario}, \ref{fig:cond_gen_mario} and in Appendix \ref{app:add_results}.
The difference in performance between CCVAE and VAEL lies in the fact that for each label there are many possible correct images. For example, in the Mario dataset, there are $6$ possible pairs of agent's positions that correspond to the label \texttt{left}. Our probabilistic logic program explicitly encodes the digits value or the single agent's positions in its probabilistic facts, and uses the variable $z_{sym}$ to compute their probabilities. On the contrary, CCVAE is not able to learn the proper mapping from the digits value or the agent's positions to the label, but it can  learn to encode only the label in the latent space $z_{sym}$.

\textbf{Conditional Image Generation.} In this task, we want to evaluate also the conditional generation ability of our approach. In Figures \ref{fig:cond_gen_digits} and \ref{fig:cond_gen_mario} we report some qualitative results for both VAEL  and CCVAE (additional results can be found in Appendix \ref{app:add_results}). As it can be seen in \ref{fig:cond_gen_digits}, VAEL always generates pairs of digits coherent with the evidence, showing also a variety of combinations. Conversely, some of the pairs generated by CCVAE do not sum to the desired value. Regarding \textit{Mario} dataset (Figure \ref{fig:cond_gen_mario}), VAEL generates pairs of states coherent with the evidence, and with different backgrounds that are preserved from one state to the subsequent one. On the contrary, CCVAE is not able to draw the agent in the generated images, thus failing the task. The reason lies, again, in the task complexity, that VAEL reduces by relying on its probabilistic logic program.

\begin{table}[htp]
\caption{Reconstructive, predictive and generative ability of VAEL and CCVAE. We use repeated trials to evaluate both the models on a test set of $10K$ images for \textit{2digit MNIST} dataset and $1344$ images for \textit{Mario} dataset.}
    \vspace{10pt}
\begin{center}
\begin{tabular}{lllllc}
\hline
\hline
\textbf{Dataset} & \textbf{Model} &  $m_{REC } (\downarrow)$ & $m_{CLASS } (\uparrow)$ & $m_{GEN } (\uparrow)$\\ 
    \hline
    \hline
    \multirow{2}{*}{\textit{2digit MNIST}}&\textit{CCVAE} & $ 1549 \pm 2$ & $0.5284 \pm 0.0051$ & $0.5143 \pm 0.0157$\\ \cline{2-5}
        & \textit{VAEL} & $\mathbf{1542} \pm \mathbf{3} $ & $\mathbf{0.8477} \pm \mathbf{0.0178}$ & $ 	\mathbf{0.7922} \pm \mathbf{0.0350}$\\ 
\hline

    \multirow{2}{*}{\textit{Mario}}&\textit{CCVAE} & $43461 \pm 209$ & $\mathbf{1.0} \pm \mathbf{0.0}$ & $0.0 \pm 0.0$\\ \cline{2-5}
        & \textit{VAEL} & $\mathbf{42734} \pm \mathbf{246} $ & $\mathbf{0.977} \pm \mathbf{0.0585}$ & $ 	\mathbf{0.8135} \pm \mathbf{0.2979}$\\ 
        \hline
\end{tabular}
\end{center}
\label{tab:base_task}
\end{table}

\begin{figure}[t]
     \centering
     \begin{subfigure}[b]{0.5\textwidth}
         \centering
         \includegraphics[width=0.8\textwidth]{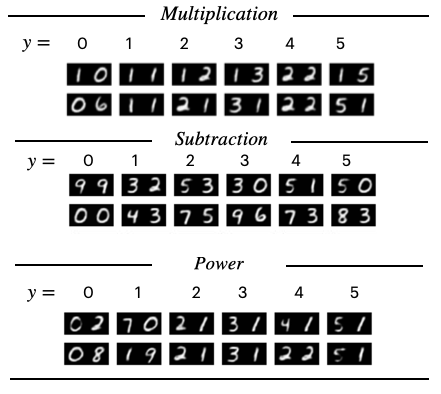}
         \caption{}
         \label{fig:task_gen_digits}
     \end{subfigure}%
     \begin{subfigure}[b]{0.5\textwidth}
         \centering
         \includegraphics[width=1.\textwidth]{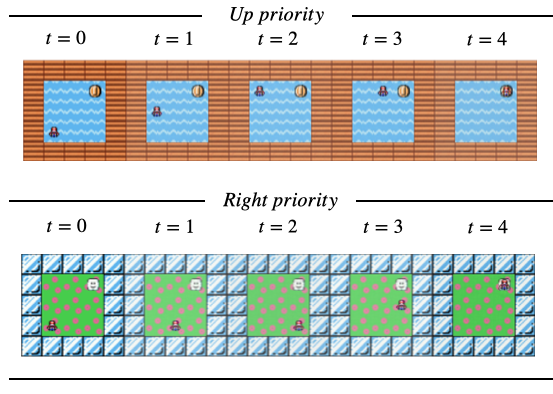}
         \caption{}
         \label{fig:task_gen_mario}
     \end{subfigure}%
  \caption{Examples of the generation ability of VAEL in  previously unseen tasks for \textit{2digit MNIST} dataset (a) and \textit{Mario} dataset (b).}

     \label{fig:task_gen}

\end{figure}

\textbf{Task Generalization.} We define several novel tasks to evaluate the task generative ability of VAEL. For \textit{2digit MNIST} dataset, we introduce the \textit{multiplication}, \textit{subtraction} and \textit{power} between two digits, while for \textit{Mario} dataset we define two shortest paths (\textit{up priority}, i.e. \texttt{up} always first, and one with \textit{right priority}, i.e. \texttt{right} always first).
To the best of our knowledge, such a level of task generalization cannot be achieved by any existing VAE framework. On the contrary, in VAEL, we can generalize by simply substituting the ProbLog program used for the training task with the program for the desired target task, without re-training the model. In Figure \ref{fig:task_gen}, we report qualitative results: in \ref{fig:task_gen_digits}, the generation is conditioned on a different label $y$ referring to the corresponding mathematical operation between the first and second digit; in \ref{fig:task_gen_mario}, the model is asked to generate a trajectory starting from the initial image ($t=0$) and following the shortest path using an \textit{up priority} or a \textit{right priority}.

In all the novel tasks of \textit{2digit MNIST} dataset (Figure \ref{fig:task_gen_digits}), VAEL generates pairs of numbers consistent with the evidence, and it also shows a variety of digits combinations by relying on the probabilistic engine of ProbLog.
This should not surprise. In fact, in all these tasks, the decoder takes as input a possible world, i.e., a specific configuration of the two digits. Therefore, the decoder is agnostic to the specific operation, which is entirely handled by the symbolic program.  For this reason, VAEL can be seamlessly applied to all those tasks that require the manipulation of two digits. The same reasoning can be extended to \textit{Mario} novel tasks (Figure \ref{fig:task_gen_mario}), where VAEL generates subsequent states consistent with the shortest path, while preserving the background of the initial state ($t=0$) thanks to the clear separation between the subsymbolic and symbolic latent components. Additional results can be found in Appendix \ref{app:add_results}. 

\textbf{Data Efficiency.}
In this task, we want to verify whether the use of a logic-based prior helps the learning in contexts characterized by data scarcity. To this goal, we define different training splits of increasing size for the \textit{addition} task of \textit{2digit MNIST} dataset. In particular, the different splits range from  $10$ up to $100$ images per pair of digits. The results (Figure \ref{fig:data_eff_graphs} in Appendix \ref{app:3mnist}) show that VAEL outperforms the baseline for all the tested sizes. In fact, with only $10$ images per pair, VAEL already performs better than CCVAE trained with $100$ images per pair. When considering $10$ images per pair, the discriminative and generative accuracies of VAEL are $0.445 \pm	0.057$ and $0.415 \pm0.0418$, whereas CCVAE trained on $100$ images per pair has a discriminative and generative accuracy of $0.121 \pm	0.006$ and $0.284 \pm 0.006$ respectively. The reason behind this disparity is that the logic-based prior helps the neural model in properly structuring the latent representation, so that one part can easily focus on recognizing individual digits and the other on capturing the remaining information in the scene. Conversely, CCVAE needs to learn how to correctly model very different pairs that sum up to the same value. 
We further investigate the performance gap between CCVAE and VAEL by running an identical experiment in a simplified dataset with only three possible digits values: $0$, $1$ and $2$. The goal is to train CCVAE on a much larger number of images per pair, which is impractical in the $10$-digits setting, due to the combinatorial nature of the task. Additional details can be found in Appendix \ref{app:3mnist}.

%% file: src/related.tex
\label{sec:related}

\textbf{Controlled image generation}. We distinguish between generative models based on text descriptions and generative models based on scene graphs. Regarding the first category, substantial effort has been devoted to devising strategies able to generate images with control (i) on object properties/attributes (e.g. shape, color, texture of objects)~\cite{reed2016generative,reed2016learning,zhang2017stackgan,zhang2019stackpp,du2020compositional}, (ii) on spatial relations between multiple objects (e.g. object A is below object B)~\cite{mansimov2016generating,oord2016contitional,hong2018inferring,liu2021learning}, (iii) or both~\cite{ramesh2021zero}. Our framework is related to these works as considering the problem of generation in a relational setting. Differently from them, we use probabilistic logic programming to encode first-order logical knowledge and to perform reasoning over this knowledge. This comes with the advantage that we can generalize to out-of-distribution relations, which consists of both the composition of previously seen relations (e.g. the multiplication can be composed by using the sum in the domain of natural numbers) and new relations (e.g. the subtraction cannot be composed by using the sum in the domain of natural numbers). 
Regarding the second category, scene graphs are used as an alternative to text descriptions to explicitly encode relations, such as spatial relations between objects~\cite{johnson2018image,oron2019scene,jiuxiang2019scene,li2019paste,mittal2019interactive,herzig2020learning,hua2021relation,chang2021survey}.
While related, our approach differs from these last as logical programs are more expressive and allow a more general reasoning than scene graphs alone.

\textbf{Unsupervised scene decomposition} We distinguish between object-oriented, part-oriented and hierarchical approaches.
The first category attempts to learn individual object representations in an unsupervised manner and to reconstruct the original image or the subsequent frame (in the case of sequential data) from these representations. Several approaches have been proposed, based on scene-mixtures~\cite{greff2017neural,sjoerd2018relational,burgess2019monet,greff2019iterative,engelcke2020genesis,locatello2020object,kipf2021conditional,stanic2021hierarchical}, spatial attention models ~\cite{gregor2015draw,eslami2016attend,crawford2019spatially} and their corresponding combination~\cite{lin2020space,jiang2020scalor}.  In the second category, a scene with an object is decomposed into its constituent parts. Specifically, an encoder and a decoder are used to decompose an object into its primitives and to recombine them to reconstruct the original object, respectively. Several approaches have been proposed for generating 3D shapes~\cite{tulsiani2017shape,li2017grass,zhu2018scores,huang2020part,kania2020ucsg,deng2020convex} and for inferring the compositional structure of the objects together with their physical interactions in videos~\cite{xu2019discovery,li2020causal,gopa2021keypoint}. These approaches focus on learning the part-whole relationships of object either by using pre-segmented parts or by using motion cues.
Last but not least, there has been recent effort focusing on integrating the previous two categories, thus learning to decompose a scene into both its objects and their respective parts, the so called hierarchical decomposition~\cite{sabour2021capsules,deng2021generative}. Our work differs in several aspects and can be considered as an orthogonal direction. First of all, we consider static images and therefore we do not exploit temporal information. Secondly, we do not provide any information about the location of the objects or their parts and use a plain autoencoder architecture to discover the objects. Therefore, we could exploit architectural advances in unsupervised scene decomposition to further enhance our framework. However, this integration is left to future investigation. Finally, our model discovers objects in a scene, by leveraging the high-level logical relations among them.

\textbf{Neuro-symbolic generation}. This is an emerging area of machine learning as demonstrated by works appeared in the last few years. For example, \cite{jiang2021generative} proposes a generative model based on a two-layered latent representation. In particular, the model introduces a global sub-symbolic latent variable, capturing all the information about a scene and a symbolic latent representation, encoding the presence of an object, its position, depth and appearance. However, the model is limited in the form of reasoning, as able to generate images with objects fulfilling only specific spatial relations. In contrast, our model can leverage a logical reasoning framework and solve tasks requiring to manipulate knowledge to answer new generative queries. 

There are two recent attempts focusing on integrating generative models with probabilistic programming~\cite{feinman2020generating,gothoskar2021program}, where reasoning is limited to spatial relationships of (parts of) the image. Moreover, \cite{gothoskar2021program} is a clear example of the difficulty of integration  the symbolic and the perceptual module. In contrast, our work provides a unified model which can  learn to generate images while perform logical reasoning at the same time.

To the best of our knowledge, the work in~\cite{kersting2020pae} represents the first attempt to integrate a generative approach with a logical framework. However, the work differs from ours in several aspects. Firstly, the authors propose a model for an image completion problem on MNIST and it is unclear how the model can be used in our learning setting and for generating images in the presence of unseen queries. Secondly, the authors propose to use sum-product networks as an interface between the logical and the neural network modules. In contrast, we provide a probabilistic graphical model which compactly integrates the two modules without requiring any additional network. Thirdly, we are the first to provide experiments supporting the benefits of such integration both in terms of task generalization and data efficiency.

\textbf{Structured priors for latent variable models}. Several structured priors have been proposed in the context of latent variable models. For example, The work in~\cite{tomczak2018vae} focuses on learning priors based on mixture distributions.~\cite{bauer2019resampled} uses rejection sampling with a learnable acceptance function to construct a complex prior. The works of~\cite{sonderby2016ladder,maaloe2019biva,vahdat2021nvae,klushyn2019learning} consider learning hierarchical priors,~\cite{chen2017variational,oord2017neural,razavi2019generating}  introduce autoregressive priors~\cite{chen2017variational}. While structured priors offer the possibility of learning flexible generative models and avoid the local minima phenomenon observed in traditional VAEs, they are quite different from ours. Indeed, our prior disentangles the latent variables to support logical reasoning. Furthermore, the structure of the logic program is interpretable.

%% file: src/appendices.tex
\appendix
\onecolumn
\section{ELBO derivation}
\label{app:elbo_derivation}
To derive the ELBO defined in (\ref{eq:loss}) we start from the maximization of the log-likelihood of the input image $x$ and the class $y$, namely
\begin{equation}
\label{eq:1A}
\log(p(x,y)) = \log\left(\int p(x,y|\mathbf{z})d\mathbf{z}\right).
\end{equation}
Recalling the generative network factorization (\ref{eq:generative_network}), we can write
\begin{equation}
\label{eq:2A}
\log(p(x,y)) = \log\left(\int p_\theta(x|z,z_{sym})p_\theta(y|z_{sym})p(z)p(z_{sym}) dz dz_{sym}\right)
\end{equation}
Then, by introducing the variational approximation $q_{\phi}(\mathbf{z}|x)$ to the intractable posterior $p_\theta(\mathbf{z}|x)$ and applying the factorization, we get
\begin{equation}
\label{eq:3A}
\log(p(x,y)) = \log\left(\int{
\frac{q_{\phi}(z|x)q_{\phi}(z_{sym}|x)}{q_{\phi}(z|x)q_{\phi}(z_{sym}|x)}
p_\theta(x|z,z_{sym}) p_\theta(y|z_{sym}) p(z) p(z_{sym}) dz dz_{sym}}\right).
\end{equation}
We now apply the \textit{Jensen’s inequality} to equation (\ref{eq:3A}) and we obtain the lower bound for the log-likelihood of $x$ and $y$ given by
\begin{equation}
\label{eq:4A}
\int{{q_{\phi}(z|x)q_{\phi}(z_{sym}|x)}\log\left(p_\theta(x|z,z_{sym}) p_\theta(y|z_{sym}) \frac{p(z) p(z_{sym})}{q_{\phi}(z|x)q_{\phi}(z_{sym}|x)} dz dz_{sym}\right).}
\end{equation}
Finally, by relying on the linearity of expectation and on logarithm properties, we can rewrite equation (\ref{eq:4A}) as
\begin{equation*}
    \mathbb{E}_{\mathbf{z}\sim q_{\phi}(\mathbf{z}|x)}\left[ \log(p_\theta(x|\mathbf{z}))\right] + \mathbb{E}_{z_{sym}\sim q_{\phi}(z_{sym}|x)}\left[\log(p_\theta(y|z_{sym})) \right]+ \mathbb{E}_{\mathbf{z} \sim q_{\phi}(\mathbf{z}|x)}\left[\log \left(\frac{p(\mathbf{z})}{q_{\phi}(\mathbf{z}|x)}\right)\right].
\end{equation*}
The last term is the negative Kullback-Leibler divergence between the variational approximation $q_{\phi}(\mathbf{z}|x)$  and the prior $p(\mathbf{z})$. This leads us to the ELBO of equation (\ref{eq:loss}), that is
\begin{align*}
    \log(p(x,y)) & \geq \mathbb{E}_{\mathbf{z} \sim q_{\phi}(\mathbf{z}|x)}\left[
    \log(p_\theta(x|\mathbf{z}))\right] +
    \mathbb{E}_{z_{sym} \sim q_{\phi}(z_{sym}|x)}\left[\log(p_\theta(y|z_ {sym})) \right] -
    \mathcal{D_{KL}}[q_{\phi}(\mathbf{z}|x)||p(\mathbf{z})]\\
    & := \mathcal{L}(\theta,\phi).
\end{align*}

\begin{figure}[htp]
\centering
    \includegraphics[width = 0.5 \textwidth]{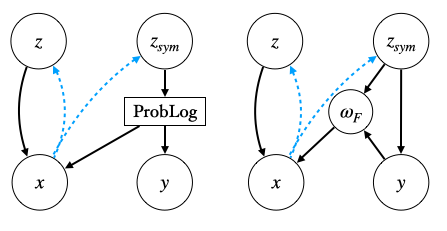}
    \caption{PGM with (left) and without (right) ProbLog box.}
     \label{fig:comparison}
\end{figure}

In VAEL graphical model (Figure \ref{fig:pgm_VAEL}),  we omit $\omega_F$ since we exploit an equivalence relation between the probabilistic graphical models (PGMs) shown in Figure \ref{fig:comparison}. Indeed, the objective for the PGM where $\omega_F$ is explicit is equivalent to the one reported in the paper. This is supported by the derivation of $\log p(x,y)$  (Eq. \ref{eq:equivalent_derivation}), which is equivalent to Eq. (\ref{eq:loss}) in our paper, where the expectation over $\omega_F$ is estimated through Gumbel-Softmax.
\begin{align}
\log p(x,y)
&=\log\int_{z,z_{sym},\omega_F} q(z,z_{sym}|x) p(x|z,\omega_F)p(y|z_{sym})p(\omega_F|z_{sym},y)\frac{p(z,z_{sym})}{q(z,z_{sym}|x)} \nonumber \\
        &\geq \int_{z,z_{sym},\omega_F} q(z,z_{sym}|x) p(\omega_F|z_{sym},y)\log p(x|z,\omega_F)p(y|z_{sym})\frac{p(z,z_{sym})}{q(z,z_{sym}|x)} \nonumber\\
        &=\mathbb{E}_{z,z_{sym},\omega_F}[\log p(x|z,\omega_F)] + \mathbb{E}_{z_{sym}}[\log p(y|z_{sym})] - KL[q(z,z_{sym}|x)\|p(z,z_{sym})] 
\label{eq:equivalent_derivation}
\end{align}

\section{ELBO estimation and Learning}
\label{app:learning}
We estimate the ELBO and its gradients w.r.t. the model parameters using standard Monte Carlo estimates of expectations \cite{kingma2013autoencoding}. 
Since both $q_{\phi}(\mathbf{z}|x)$ and $p(\mathbf{z})$ are chosen to be Gaussian distributions, the Kullback-Leibler divergence in (\ref{eq:loss}) can be integrated analytically by relying on its closed form. Thus, only the expected reconstruction and query errors $ \mathcal{L}_{REC}( \theta,\phi)$ and $\mathcal{L}_{Q}( \theta,\phi)$ require estimation by sampling.\\
We can therefore define the ELBO estimator as
\begin{equation}
  \mathcal{L}(\theta,\phi) \approx \mathcal{\tilde{L}}(\theta,\phi; \epsilon) = \mathcal{\tilde{L}}_{REC}( \theta,\phi;\mathbb{\epsilon}) + \mathcal{\tilde{L}}_{Q}( \theta,\phi;\mathbb{\epsilon}) - \mathcal{D_{KL}}[q_{\phi}(\mathbf{z}|x)||p(\mathbf{z})].
  \label{eq:loss_est}
\end{equation}
The estimators of $\mathcal{L}_{REC}$ and $\mathcal{L}_{Q}$ can be written as
\begin{align}
    \tilde{\mathcal{L}}_{REC}( \theta,\phi;\mathbb{\epsilon}) &= {{1}\over{N} }\sum_{n =1}^{N}(\log(p_\theta(x|\hat{\mathbf{z}}^{(n)}))) \\
    \mathcal{\tilde{L}}_{Q}( \theta,\phi;\mathbb{\epsilon}) &= {{1}\over{N} }\sum_{n =1}^{N}(\log(p_\theta(y|\hat{z}_{sym}^{(n)})))
\end{align}
where
\begin{align*}
    \hat{\mathbf{z}}^{(n)} &= \{\hat{z}^{(n)}, \hat{z}_{sym}^{(n)} \} := \mathbb{\mu}(x) + \mathbb{\sigma}(x) \mathbb{\epsilon}^{(n)}, \\
    \mathbb{\epsilon}^{(n)} &\sim \mathcal{N}(0,1).
\end{align*}

During the training, we aim at maximizing $\mathcal{L}( \theta,\phi)$ with respect to both the encoder and the decoder parameters, we therefore need to compute the gradient w.r.t. $\theta$ and $\phi$.
Since any sampling operation prevents back-propagation, we need to reparametrize the two sampled variables $\mathbf{z}$ and $\omega$.
Due to their nature, we use the well-known \textit{Reparametrization Trick} \cite{kingma2013autoencoding} for the Gaussian $\mathbf{z}$, while we exploit the \textit{Categorical Reparametrization with Gumbel-Softmax } \cite{jang2017categorical} for the discrete variable $\omega$ corresponding to the sampled possible world.\\ In particular, by defining $\omega$ as the one-hot encoding of the possible worlds, we have
\begin{equation}
\hat{\omega}_i= {{\exp((\log \pi_i + \hat{g}_i)/\lambda}\over{\sum_{j =1}^J\exp((\log \pi_j + \hat{g}_j)/\lambda)}}, \text{ with } \hat{g}_i \sim Gumbel(0,1) 
\label{eq:gumbel}
\end{equation}
where $J$ is the number of possible worlds (e.g. all the possible pairs of digits), and $\pi_i$ depends on $z^i_{sym}$, which is reparametrized with the Gaussian Reparametrization Trick.
In Algorithm \ref{training_alg} we report VAEL training algorithm .

        \begin{algorithm}[H]
    \caption{VAEL Training.}
    \label{training_alg}

    \KwData{Set of images $\mathcal{X}$}
    $\theta, \phi \leftarrow$ Initialization of paramters\\
    \Repeat{convergence of parameters $(\theta, \phi)$}{
    \BlankLine
    \emph{Forward Phase}\\
    $x \leftarrow$ Training sample\\
    $\mathbf{z} = [z, z_{sym}] \sim q(\mathbf{z} \mid x)$\\
    $p = MLP(z_{sym})$\\
    $\omega_F  \sim P(\omega_F; p)$\\
    $y \sim P(y; p)$ \\
    $\tilde{x} \sim  p(x | z, \omega_F) $\\
    \BlankLine
    \emph{Backward Phase}\\
    $\mathbf{g} \leftarrow \nabla_{\theta, \phi}\mathcal{L}(\theta,\phi)$ \\
    $\theta, \phi \leftarrow$ 
    Update parameters using gradients $\mathbf{g}$\\
     }
\end{algorithm}

\section{Additional supervision for MNIST Task Generalization}
\label{app:add_sup}

During the training on \textit{2digit MNIST} dataset, the model may learn a mapping between symbol and meaning that is logically correct, but different from the desired one. Indeed, the two symbols $1$ and $2$ used for the left and right positions, respectively, of a handwritten digit in an image are just an assumption. However, VAEL may switch the pairs $(3,2)$ and $(2,3)$, since they both sum up to $5$. This would prevent VAEL from generalizing to tasks involving non-commutative operations (i.e. \textit{subtraction} and \textit{power}).

To solve this issue, we simply introduce additional supervision on the digits of very few images ($1$ image per pair of digits, i.e. $100$ images in total) to guide the model toward the desired symbols interpretation. This has to be intended just as an agreement between the model and the human. To include this supervision in the training procedure, we add a regularizer term to the ELBO defined in (\ref{eq:loss}), namely

\begin{equation}
\mathcal{L_{SUP}}(\theta,\phi) := \mathcal{L}(\theta,\phi) + \mathcal{L}_{digits}( \theta,\phi) \label{eq:sup_loss}
\end{equation}

where
\begin{equation}
\mathcal{L}_{digits}(\theta,\phi) = 
     \mathbb{E}_{z_{sym}\sim q_{\phi}(z_{sym}|x)}[
       log(p_\theta(y_{digits}|z_{sym})]].
\label{eq:digits_loss}
\end{equation}

In equation (\ref{eq:digits_loss}), $y_{digits}$ refers to the labels over the digits (e.g. for image \imgZERONE ~ we have $ y_{digits} = [0,1]$).\\
Such a digit-level supervision can be easily done by virtue of ProbLog inference, that allows us to retrieve the predicted label of each digit in an image by relying on the query over the digits values.

\section{Implementation details}
\label{app:implem}
\subsection{VAEL}
In Tables \ref{tab:vael_mnist} and  \ref{tab:vael_mario}  we report the architectures of VAEL for \textit{2digit MNIST} and \textit{Mario} dataset. For both the datasets we performed a model selection by minimizing the objective function computed on a validation set of $12,000$ samples for \textit{2digit MNIST} and $2,016$ samples for \textit{Mario}.  In all the experiments we trained the model with Adam \cite{kingma2017adam}. The explored hyper-parameters values are reported in Section \ref{app:implem:opt}.

For \textit{2digit MNIST}, the resulting best configuration is: latent space $z \in \mathbb{R}^M$ , $z_{sym} \in \mathbb{R}^N$ with dimension $M = 8$ and $N = 15$; weights $0.1$, $1 \times 10^{-5}$ and $1.0$ for the reconstruction, Kullback-Leibler and classification term of the ELBO respectively; learning rate $1 \times 10 ^{-3}$. 

For \textit{Mario}, we obtain: latent space $z \in \mathbb{R}^M$ , $z_{sym} \in \mathbb{R}^N$ with dimension $M = 30$ and $N = 18$; weights $1 \times 10^{1}$, $1 \times 10^{1}$ and $1  \times 10^{4}$ for the reconstruction, Kullback-Leibler and classification term of the ELBO respectively;  learning rate $1 \times 10 ^{-4}$.

\begin{table}[htp]
\caption{VAEL architectures for \textit{2digit MNIST} dataset.}

\centering
\begin{tabular}{cc}
\hline $\text { \textbf{Encoder} }$ & $\text { \textbf{Decoder} }$ \\
\hline $\text { Input } 28 \times 56 \times 1 \text { channel image }$ & $\text { Input } \in \mathbb{R}^{M+20} $\\
$64 \times 1 \times 4 \times 4 \text { Conv2d stride } 2 \text { \& ReLU }$ & $(M+20) \times 256 \text { Linear layer }$ \\
$128 \times 64 \times 4 \times 4 \text { Conv2d stride } 2 \text { \& ReLU }$ & $256 \times 128 \times 5 \times 4 \text { ConvTranspose2d stride } 2 \text { \& ReLU } $\\
$256 \times 128 \times 4 \times 4 \text { Conv2d stride } 2 \text { \&ReLU }$ & $128 \times 64 \times 4 \times 4 \text{ ConvTranspose2d stride } 2  \text { \& ReLU }$ \\
$256 \times 2 (M+N) \text { Linear layer }$ & $1 \times 64 \times 4 \times 4 \text { ConvTranspose2d stride 2 \& Sigmoid }$\\
\hline
\end{tabular}

\vspace{10pt}

\begin{tabular}{c}
\hline \text { \textbf{MLP \& ProbLog} } \\
\hline             
 $ \text { Input } \in \mathbb{R}^{N}   $ \\
             $N \times 20$ \text { Linear layer } \& ReLU
            \\
           $20 \times 20$ \text { Linear layer } 
            \\
            ProbLog (IN dim: $20$, OUT dim: $100$)
            \\
\hline
\end{tabular}

\label{tab:vael_mnist}
\end{table}

\begin{table}[htp]
\caption{VAEL architectures for \textit{Mario} dataset.}

\centering
\begin{tabular}{cc}
\hline $\text { \textbf{Encoder} }$ & $ \text { \textbf{Decoder} }$ \\
\hline  $ \text { Input } 200 \times 100 \times 3 \text { channel image } $  & $  \text { Input } \in \mathbb{R}^{M+9} $  \\
 $ 64 \times 3 \times 5 \times 5 \text { Conv2d stride } 2 \text { \& SELU }  $ & $  (M+9) \times 512 \text { Linear layer } $  \\
 $ 128 \times 64 \times 5 \times 5 \text { Conv2d stride } 2 \text { \& SELU }  $ & $  512 \times 256 \times 5 \times 5 \text { ConvTranspose2d stride } 2 \text { \& SELU } $  \\
 $ 256 \times 128 \times 5 \times 5 \text { Conv2d stride } 2 \text { \& SELU }  $ &  $ 256 \times 128 \times 5 \times 5 \text{ ConvTranspose2d stride } 2 \text {\& SELU } $  \\
 $ 512 \times 256 \times 5 \times 5 \text { Conv2d stride } 2  \text { \& SELU } $  &  $ 128 \times 64 \times 5 \times 5 \text{ ConvTranspose2d stride } 2 \text {\& SELU } $  \\
 $ 512 \times 2 (M+9) \text { Linear layer } $  & $  3 \times 64 \times 5 \times 5 \text { ConvTranspose2d stride 2 \& Sigmoid } $  \\
\hline
\end{tabular}

\vspace{10pt}

\begin{tabular}{c}
\hline \text { \textbf{MLP \& ProbLog} } \\
\hline             
 $ \text { Input } \in \mathbb{R}^{N}  $  \\
             $N \times 20$ \text { Linear layer } \& ReLU
            \\
           $20 \times 9$ \text { Linear layer } 
            \\
            ProbLog (IN dim: $18$, OUT dim: $24$)
            \\
\hline
\end{tabular}

\label{tab:vael_mario}
\end{table}

\subsection{CCVAE}
In the original paper \cite{joy2021capturing}, there was a direct supervision on each single element of the latent space.
To preserve the same type of supervision in our two digits addition task, where the supervision is on the sum and not directly on the single digits, we slightly modify the encoder and decoder mapping functions of CCVAE. By doing so, we ensure the correctness of the approach without changing the graphical model.
The original encoder function learns from the input both the mean $\mu$ and the variance $\sigma$ of the latent space distribution, while the decoder gets in input the latent representation $\mathbf{z} = \{z_{sym}, z\}$ (please refer to the original paper for more details \cite{joy2021capturing}). In our modified version, the encoder only learns the variance, while the mean is set to be equal to the image label $\mu = y$, and the decoder gets in input the label directly $\mathbf{z^*} := \{y, z\}$.

In Tables \ref{tab:ccvae_mnist} and  \ref{tab:ccvae_mario}  we report the architectures of CCVAE for \textit{2digit MNIST} and \textit{Mario} dataset. For both the datasets we performed a model selection by minimizing the objective function computed on a validation set of $12,000$ samples for \textit{2digit MNIST} and $2,016$ samples for \textit{Mario}.  In all the experiments we trained the model with Adam \cite{kingma2017adam}. The explored hyper-parameters values are reported in Section \ref{app:implem:opt}.

For \textit{2digit MNIST}, the resulting best configuration is: latent space $z_{sym} \in \mathbb{R}^N$ with dimension equal to the number of classes $N = 19$ (due to the one-to-one mapping between $z_{sym}$ and the label $y$);  latent space $z \in \mathbb{R}^M$ with dimension $M = 8$, model objective reconstruction term with weight $0.05$, while the other ELBO terms with unitary weights; learning rate $1 \times 10 ^{-4}$. 

For \textit{Mario}, we obtain: latent space $z_{sym} \in \mathbb{R}^N$ with dimension equal to the number of classes $N = 4$; latent space $z \in \mathbb{R}^M$ with dimension $M = 300$, model objective Kullback-Leibler term and classification term with weight $1 \times 10^4$ and $1 \times 10^3$ respectively, while the other ELBO terms with unitary weights; learning rate $1 \times 10 ^{-4}$.

\begin{table}[htp]
\caption{CCVAE architectures for \textit{2digit MNIST} dataset.}
\begin{tabular}{cc}
\hline  $ \text { \textbf{Encoder} }  $ & $  \text { \textbf{Decoder} }  $ \\
\hline  $ \text { Input } 28 \times 56 \times 1 \text { channel image } $  &  $ \text { Input } \in \mathbb{R}^{M+N}  $ \\
 $ 64 \times 1 \times 4 \times 4 \text { Conv2d stride } 2 \text { \& ReLU }  $ & ( $ M+N) \times 256 \text { Linear layer }  $ \\
 $ 128 \times 64 \times 4 \times 4 \text { Conv2d stride } 2 \text { \& ReLU }  $ &  $ 256 \times 128 \times 5 \times 4 \text { ConvTranspose2d stride } 2 \text { \& ReLU }  $ \\
 $ 256 \times 128 \times 4 \times 4 \text { Conv2d stride } 2 \text { \&ReLU } $  &  $ 128 \times 64 \times 4 \times 4 \text{ ConvTranspose2d stride } 2 \& \text { ReLU } $  \\
 $ 256 \times 2 (M+N) \text { Linear layer }  $ &  $ 1 \times 64 \times 4 \times 4 \text { ConvTranspose2d stride 2 \& Sigmoid } $  \\
\hline
\end{tabular}
\label{tab:ccvae_mnist}
\end{table}

\begin{table}[htp]
\caption{CCVAE architectures for \textit{Mario} dataset.}
\begin{tabular}{cc}
\hline  $ \text { \textbf{Encoder} } $  & $  \text { \textbf{Decoder} }  $ \\
\hline  $ \text { Input } 200 \times 100 \times 3 \text { channel image }  $ & $  \text { Input } \in \mathbb{R}^{M+N} $  \\
 $ 64 \times 3 \times 5 \times 5 \text { Conv2d stride } 2 \text { \& SELU }  $ &  $ (M+N) \times 512 \text { Linear layer } $  \\
 $ 128 \times 64 \times 5 \times 5 \text { Conv2d stride } 2 \text { \& SELU }  $ & $  512 \times 256 \times 5 \times 5 \text { ConvTranspose2d stride } 2 \text { \& SELU } $  \\
 $ 256 \times 128 \times 5 \times 5 \text { Conv2d stride } 2 \text { \&SELU }  $ &  $ 256 \times 128 \times 5 \times 5 \text{ ConvTranspose2d stride } 2 \& \text { SELU }  $ \\
 $ 512 \times 256 \times 5 \times 5 \text { Conv2d stride } 2  \text { \&SELU }  $ & $  128 \times 64 \times 5 \times 5 \text{ ConvTranspose2d stride } 2 \& \text { SELU }  $ \\
 $ 512 \times 2 (M+N) \text { Linear layer }  $ & $  3 \times 64 \times 5 \times 5 \text { ConvTranspose2d stride 2 \& Sigmoid } $  \\
\hline
\end{tabular}
\label{tab:ccvae_mario}
\end{table}

\subsection{Classifiers}
\label{app:class_mnist}

In Table \ref{tab:mnist_class} we report the architecture of the classifier used to measure the generative ability of VAEL and CCVAE for \textit{2digit MNIST} dataset. We trained the classifier on $60,000$ MNIST images \cite{lecun1998mnist} for $15$ epochs with SGD with a learning rate of $1 \times 10^{-2}$ and a momentum of $0.5$, achieving $0.97$ accuracy on the test set.
\begin{table}[htp]
\centering
\caption{}
\begin{tabular}{l}
\hline
\multicolumn{1}{c}{\textbf{MNIST classifier} (\textit{2digit MNIST})} \\ \hline
Input $28\times 28 \times 1$ channel image                   \\
Linear layer $784 \times 128$  \& ReLU                \\
Linear layer $128 \times 64$  \& ReLU                 \\
 Linear layer  $64 \times 10$ \& LogSoftmax             \\ \hline
\end{tabular}

\label{tab:mnist_class}
\end{table} 

In Table \ref{tab:mario_class} we report the architecture of the classifier used to measure the generative ability of VAEL and CCVAE for \textit{Mario} dataset. We trained the classifier on $9,140$ single state images of \textit{Mario} dataset for $10$ epochs with Adam \cite{kingma2017adam} optimizer with a learning rate of $1 \times 10^{-4}$, achieving $1.0$ accuracy on the test set.

\begin{table}[htp]
\centering
\caption{}
\begin{tabular}{l}
\hline
\multicolumn{1}{c}{\textbf{MNIST classifier} (\textit{Mario})} \\ \hline
Input $100\times 100 \times 3$ channels image                   \\
Conv layer $5 \times 5 \times 32$ \& SELU \\
Conv layer $5 \times 5 \times 64$ \& SELU \\
Conv layer $5 \times 5 \times 128$ \& SELU \\
 Linear layer $2048 \times 9$\\ \hline
\end{tabular}
\label{tab:mario_class}
\end{table} 

\subsection{Optimization} \label{app:implem:opt}

Experiments are conducted on a single Nvidia GeForce 2080ti 11 GB.  Training consumed $\sim 2 GB$
for \textit{2digit MNIST} dataset and $\sim 2.8 GB$ for \textit{Mario} dataset, taking around $1$ hour and $15$ minutes to complete $100$ epochs for \textit{2digit MNIST}  and $1$ hour and $30$ minutes to complete $100$ epochs for \textit{Mario} dataset. As introduced in the previous sections, we performed a model selection based on ELBO minimization for both the model. 

In the following bullet lists, $lr$ refers to the learning rate, $z, z_{sym}$ refer to the latent vectors dimensions, $W_{REC}, W_{KL}, W_{Q}$ refer to the weights of $\mathcal{L}_{REC}, \mathcal{D}_{KL}, \mathcal{L}_{Q}$ terms of VAEL objective function, and $W_{REC}, W_{KL}, W_{q(y\mid z_{sym})}, W_{q(y\mid x)}$ refer to the corresponding terms of CCVAE objective function (please refer to the original paper for more details \cite{joy2021capturing}). 

For \textit{2digit MNIST} we explore the following values; we repeat the model training  $5$ times for each configuration.
\begin{itemize}
    \item VAEL
    \begin{itemize}
        		 \item $z\in \{8, 9, 10\}$
		 \item $z_{sym} \in \{15, 19\}$
		 \item $lr \in \{0.0001, 0.001\}$
		 \item $W_{REC} \in \{0.0001, 0.001, 0.01, 0.1, 1, 10, 100\}$
		 \item $W_{KL}  \in \{0.00001, 0.0001, 0.001\}$
		 \item $W_{Q} \in \{1, 5\}$
    \end{itemize}
    \item CCVAE
    \begin{itemize}
		 \item $z_{sym} \in \{8, 10, 15, 20, 30\}$
		 \item $lr \in \{0.00001, 0.0001\}$
		 \item $W_{KL} \in \{0.0001, 0.001, 0.01, 0.1, 1, 10, 100\}$
		 \item $W_{REC} \in \{0.01, 0.1, 1, 10, 100\}$
		 \item $W_{q(y\mid z_{sym})} \in \{0.01, 0.1, 1, 10, 100\}$
		 \item $W_{q(y\mid x)} \in \{0.01, 0.1, 1, 10, 100\}$
    \end{itemize}
\end{itemize}

For \textit{Mario} we explore the following values; we repeat the model training  $5$ times for each configuration.
\begin{itemize}
    \item VAEL
    \begin{itemize}
        \item $z\in \{20, 25, 30, 35, 40\}$
		 \item $z_{sym} \in \{18, 20\}$
		 \item $lr \in \{0.0001, 0.0005\}$

		 \item $W_{REC} \in \{1, 10\}$
		 \item $W_{KL}\in \{0.1, 1, 10\}$
		 \item $W_{Q} \in \{1,100,10000\}$
    \end{itemize}
    \item CCVAE
    \begin{itemize}
		\item $z_{sym}\in \{3, 4, 5, 10, 20, 30, 40, 50, 100, 200, 300, 400\}$
		\item $lr \in \{0.0001,0.0005\}$
		\item $W_{KL} \in \{0.0, 0.0001, 0.001, 0.01, 0.1, 1, 10, 100, 1000\}$
		\item $W_{REC} \in \{1, 10\}$
		\item $W_{q(y\mid z_{sym})} \in \{1, 10, 100\}$
		\item $W_{q(y\mid x)} \in \{1, 10, 100, 1000\}$

    \end{itemize}
\end{itemize}

\section{Additional Results}
\label{app:add_results}

Here we report some additional results for the tasks described in Section \ref{sec:results}.

\begin{figure*}[htp]
  \begin{center}
    \includegraphics[width=1.\textwidth]{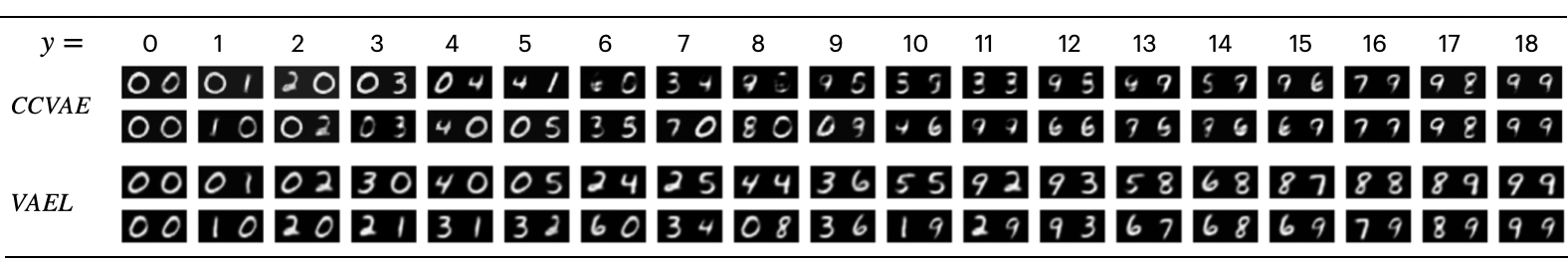}
  \end{center}
  \caption{Conditional generation for CCVAE and VAEL for \textit{2digit MNIST} dataset. In each column the generation is conditioned on a different sum $y$ between the two digits.}
\label{fig:cond_gen_extend}

\end{figure*}

\begin{figure*}[htp]
  \begin{center}
    \includegraphics[width=1.\textwidth]{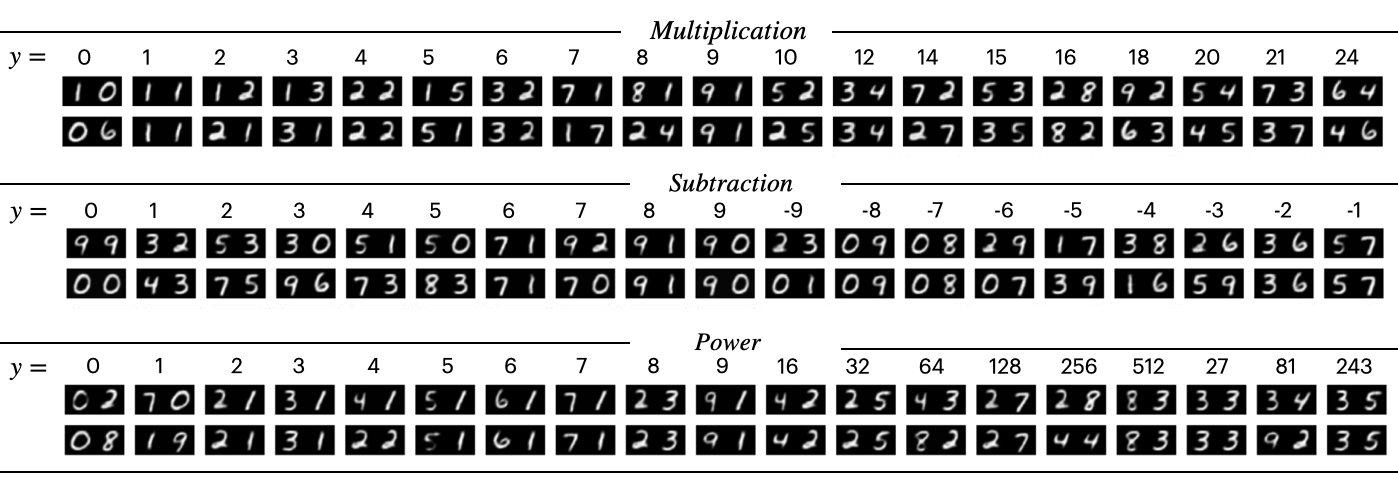}
  \end{center}
  \caption{Examples of the generation ability of VAEL in $3$ previously unseen tasks for \textit{2digit MNIST} dataset. In each column the generation is conditioned on a different label $y$ referring to the corresponding mathematical operation between the first and second digit.}
\label{fig:task_gen_extend}
\end{figure*}

Figures \ref{fig:cond_gen_extend} and \ref{fig:task_gen_extend} show additional qualitative results for the \textit{Conditional Image Generation} and \textit{Task Generalization} experiments relative to \textit{2digit MNIST} dataset. 

\begin{figure*}[htp]
  \begin{center}
    \includegraphics[width=0.83\textwidth]{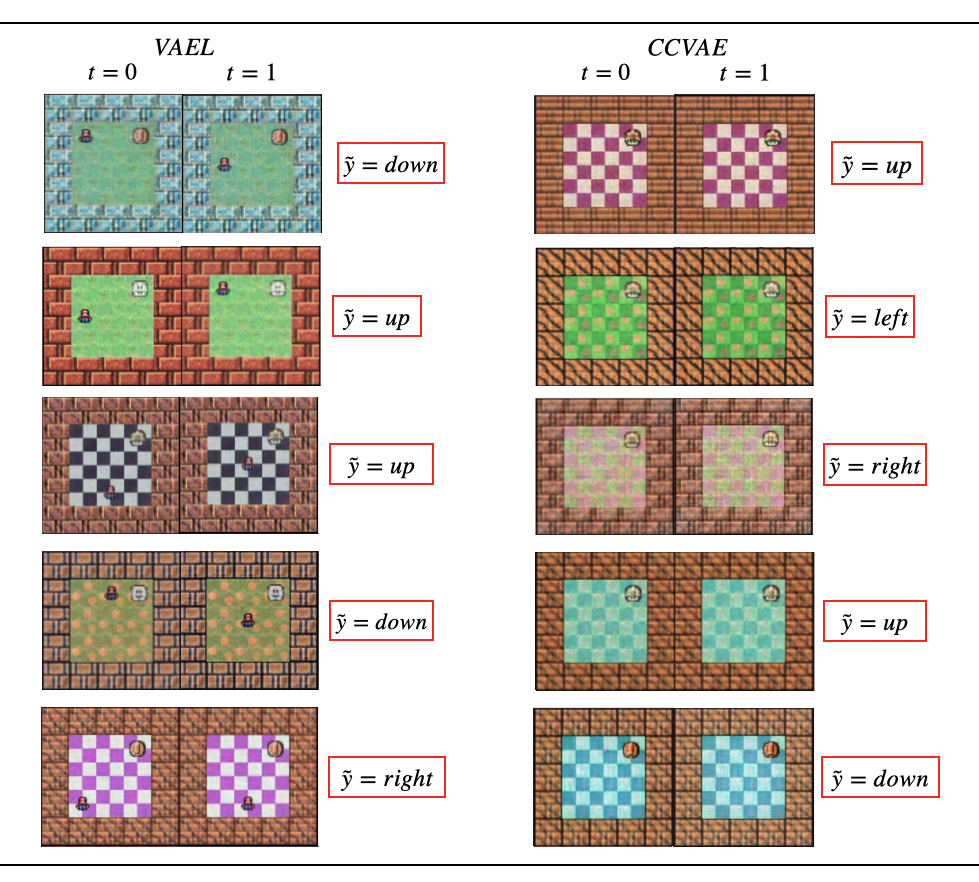}
  \end{center}
  \caption{Examples of the generation ability of CCVAE and VAEL for \textit{Mario} dataset.}
\label{fig:mario_cond_gen_extend}
\end{figure*}

\begin{figure*}[htp]
  \begin{center}\includegraphics[width=0.8\textwidth]{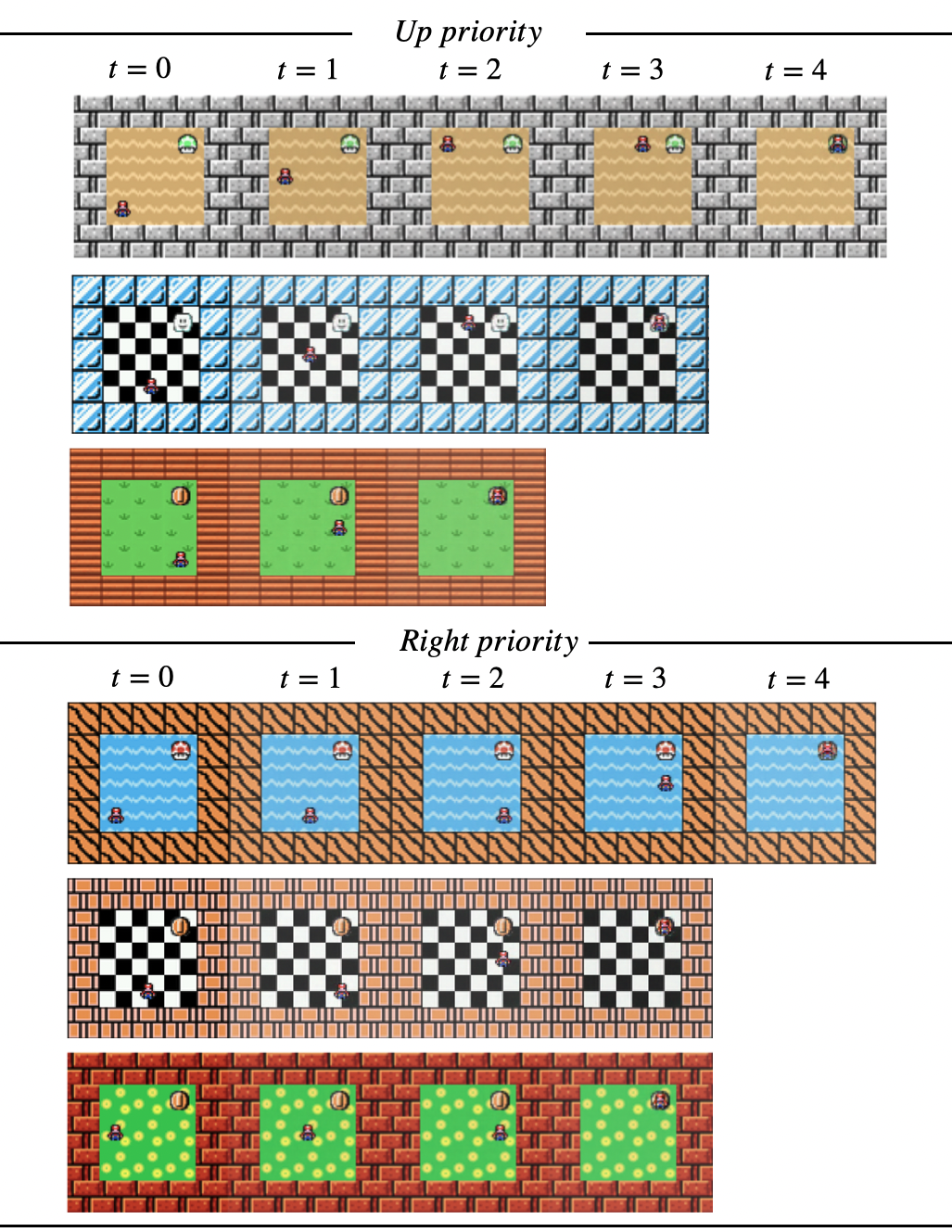}
  \end{center}
  \caption{Examples of the generation ability of VAEL in previously unseen tasks for \textit{Mario} dataset. In each row, VAEL generates a trajectory starting from the initial image (t = 0) and following the shortest path using an \textit{up priority} or a \textit{right priority}.}
\label{fig:mario_task_gen_ext}
\end{figure*}

In Figures \ref{fig:mario_cond_gen_extend} and \ref{fig:mario_task_gen_ext}, we report some additional examples of  \textit{Image Generation} and \textit{Task Generalization} for \textit{Mario} dataset. As it can be seen in Figure \ref{fig:mario_task_gen_ext}, VAEL is able to generate subsequent states consistent with the shortest path, whatever the agent's position in the initial state ($t=0$). Moreover, the model generates states that are consistent with the initial one in terms of background.

\begin{figure*}[htp]
  \begin{center}\includegraphics[width=1.\textwidth]{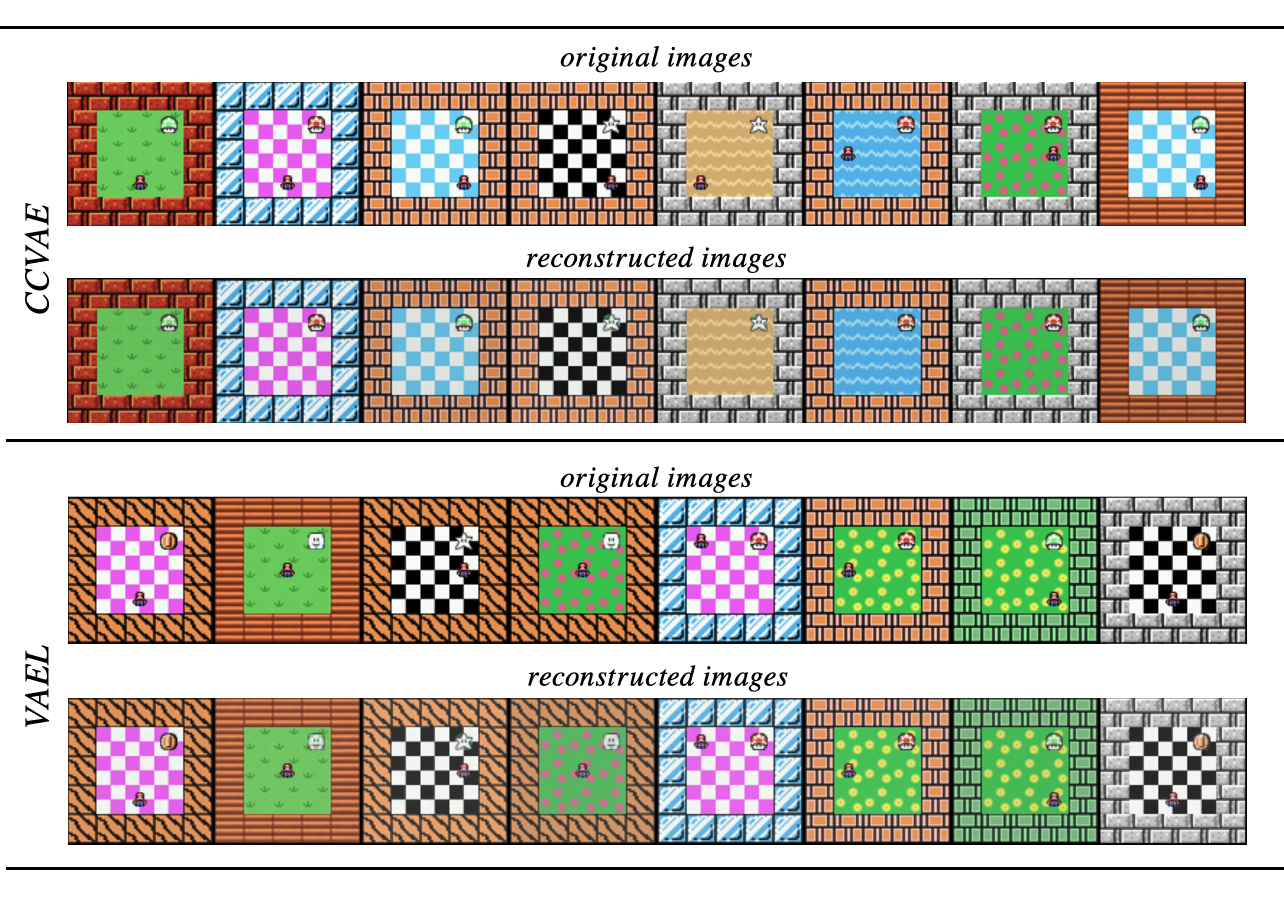}
  \end{center}
  \caption{Examples of reconstructive ability of CCVAE and VAEL trained on \textit{Mario} dataset.}
\label{fig:recon_mario}
\end{figure*}

Figure \ref{fig:recon_mario} shows some examples of image reconstruction for CCVAE. As it can be seen, CCVAE focuses only on reconstructing the background and discards the small portion of the image containing the agent, thus causing the disparity in the reconstructive and generative ability between VAEL and CCVAE (Table \ref{tab:base_task}).

\newpage
\section{Data Efficiency: simplified setting}
\label{app:3mnist}
\begin{figure*}[htp]
  \begin{center}
    \includegraphics[width=1.\textwidth]{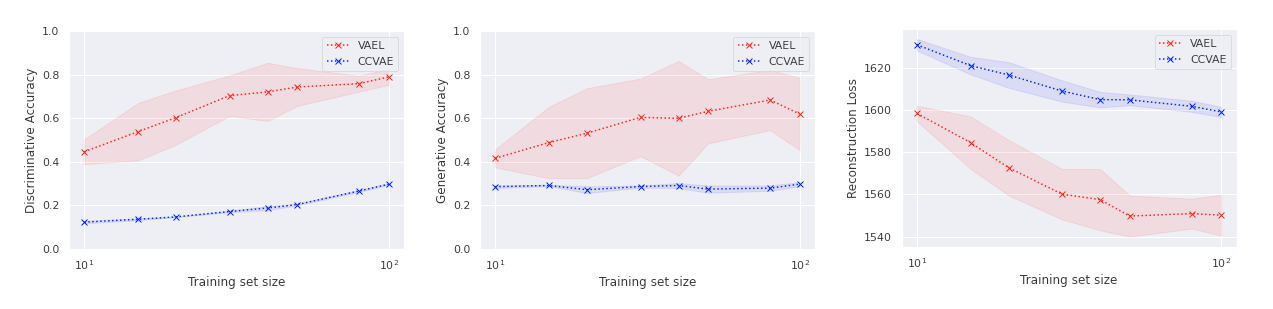}
  \end{center}
  \caption{Discriminative, generative and reconstructive ability of VAEL (red) and CCVAE (blue) trained in contexts characterized by data scarcity. Both the models are evaluated on the same test set. The training size refers to the number of samples per pair of digits see during the training.}
\label{fig:data_eff_graphs}
\end{figure*}

We compare VAEL and CCVAE discriminative, generative and reconstructive ability when varying the training size of \textit{2digit MNIST} dataset. As it can be seen in Figure \ref{fig:data_eff_graphs}, VAEL outperforms the baseline for all the tested sizes. In fact, with only $10$ images per pair VAEL already performs better than CCVAE trained with $100$ images per pair.

\begin{figure*}[htp]
  \begin{center}
    \includegraphics[width=1.\textwidth]{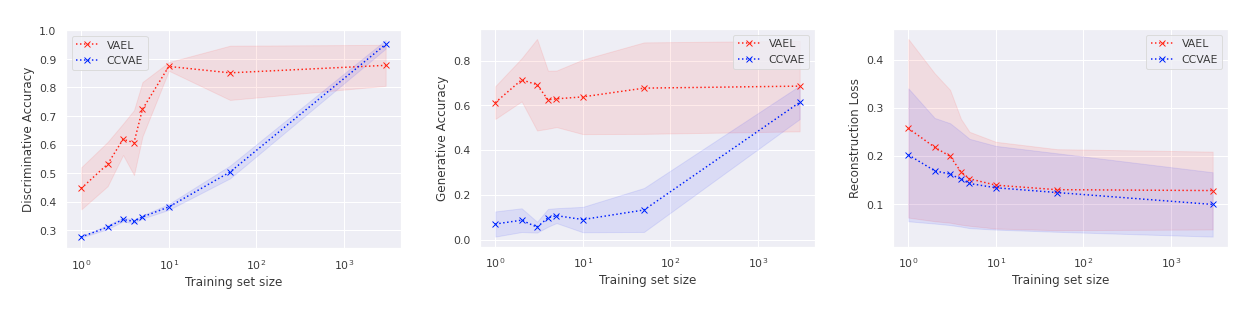}
  \end{center}
  \caption{Discriminative, generative and reconstructive ability of VAEL (red) and CCVAE (blue) trained in contexts characterized by data scarcity. Both the models are evaluated on the same test set. The training size refers to the number of samples per pair of digits see during the training.}
\label{fig:data_eff_graph_3}
\end{figure*}

To further investigate the performance gap between CCVAE and VAEL in the \textit{Data Efficiency} task \ref{sec:results}, we run an identical experiment in a simplified dataset with only three possible digits values: $0$, $1$ and $2$. The goal is to train CCVAE on a much larger number of images per pair, which is impractical in the $10$-digits setting, due to the combinatorial nature of the task. 
The dataset consists of $30,000$ images of two digits taken from the MNIST dataset \cite{lecun1998mnist}. We use 80\%, 10\%, 10\% splits for the train, validation and test sets, respectively. As for the $10$-digits dataset, each image in the dataset has dimension $28 \times 56$ and is labelled with the sum of the two digits. 
 In Figure \ref{fig:data_eff_graph_3} we compare VAEL and CCVAE discriminative, generative and reconstructive ability when varying the training size. In this simplified setting, CCVAE requires around $2500$ images per pair to reach the accuracy that VAEL achieves trained with only $10$ images per pair.

%% file: paper.bib
@inproceedings{kingma2017adam,
    author    = {Diederik P. Kingma and Jimmy Ba},
    title     = {{Adam: A Method for Stochastic Optimization}},
    booktitle = {ICLR},
    year      = {2015}
}

@inproceedings{chen2017variational,
    author    = {Xi Chen and Diederik P. Kingma and Tim Salimans and Yan Duan and Prafulla Dhariwal and John Schulman and Ilya Sutskever and Pieter Abbeel},
    title     = {{Variational Lossy Autoencoder}},
    booktitle = {ICLR},
    year      = {2017}
}

@inproceedings{bauer2019resampled,
    author    = {Matthias Bauer and Andriy Mnih},
    title     = {{Resampled Priors for Variational Autoencoders}},
    booktitle = {AISTATS},
    pages     = {66--75},
    year      = {2019}
}

@inproceedings{klushyn2019learning,
    author    = {Alexej Klushyn and Nutan Chen and Richard Kurle and Botond Cseke and Patrick van der Smagt},
    title     = {{Learning Hierarchical Priors in VAEs}},
    booktitle = {NeurIPS},
    pages     = {2866--2875},
    year      = {2019}
}

@inproceedings{tomczak2018vae,
    author    = {Jakub M. Tomczak and Max Welling},
    title     = {{VAE with a VampPrior}},
    booktitle = {AISTATS},
    pages     = {1214--1223},
    year      = {2018}
}

@article{lecun1998mnist,
    title={Gradient-based learning applied to document recognition},
    author={LeCun, Yann and Bottou, L{\'e}on and Bengio, Yoshua and Haffner, Patrick},
    journal={IEEE},
    volume={86},
    number={11},
    pages={2278--2324},
    year={1998}
}

@inproceedings{vahdat2021nvae,
    author    = {Arash Vahdat and Jan Kautz},
    title     = {{NVAE: A Deep Hierarchical Variational Autoencoder}},
    booktitle = {NeurIPS},
    year      = {2020}
}

@inproceedings{sonderby2016ladder,
    author    = {Casper Kaae S{\o}nderby and Tapani Raiko and Lars Maal{\o}e and S{\o}ren Kaae S{\o}nderby and Ole Winther},
    title     = {{Ladder Variational Autoencoders}},
    booktitle = {NeurIPS},
    pages     = {3738--3746},
    year      = {2016}
}

@inproceedings{jang2017categorical,
    author    = {Eric Jang and Shixiang Gu and Ben Poole},
    title     = {{Categorical Reparameterization with Gumbel-Softmax}},
    booktitle = {ICLR},
    year      = {2017}
}

@inproceedings{kingma2013autoencoding,
    author    = {Diederik P. Kingma and Max Welling},
    title     = {{Auto-Encoding Variational Bayes}},
    booktitle = {ICLR},
    year      = {2014}
}

@inproceedings{oord2017neural,
    author    = {A{\"{a}}ron van den Oord and Oriol Vinyals and Koray Kavukcuoglu},
    title     = {{Neural Discrete Representation Learning}},
    booktitle = {NeurIPS},
    pages     = {6306--6315},
    year      = {2017}
}

@inproceedings{razavi2019generating,
    author    = {Ali Razavi and A{\"{a}}ron van den Oord and Oriol Vinyals},
    title     = {{Generating Diverse High-Fidelity Images with VQ-VAE-2}},
    booktitle = {NeurIPS},
    pages     = {14837--14847},
    year      = {2019}
}

@inproceedings{feinman2020generating,
    author    = {Reuben Feinman and Brenden M. Lake},
    title     = {{Generating New Concepts with Hybrid Neuro-Symbolic Models}},
    booktitle = {CogSci},
    year      = {2020}
}

@inproceedings{kingma2014semisupervised,
    author    = {Diederik P. Kingma and Shakir Mohamed and Danilo Jimenez Rezende and Max Welling},
    title     = {{Semi-supervised Learning with Deep Generative Models}},
    booktitle = {NeurIPS},
    pages     = {3581--3589},
    year      = {2014},
}

@inproceedings{joy2021capturing,
    author    = {Tom Joy and Sebastian M. Schmon and Philip H. S. Torr and Siddharth Narayanaswamy and Tom Rainforth},
    title     = {{Capturing Label Characteristics in VAEs}},
    booktitle = {ICLR},
    year      = {2021}
}

@inproceedings{kersting2020pae,
    title={{Sum-Product Logic: Integrating Probabilistic Circuits into DeepProbLog}}, 
    author={A. Skryagin and K. Stelzner and A. Molina and F. Ventola and Z. Yu and K. Kersting},
    booktitle={ICML 2020 Workshop on Bridge Between Perception and
Reasoning: Graph Neural Networks and Beyond},
    year={2020}
}

@inproceedings{jiang2021generative,
    author    = {Jindong Jiang and Sungjin Ahn},
    title     = {{Generative Neurosymbolic Machines}},
    booktitle = {NeurIPS},
    year      = {2020},
}

@inproceedings{manhaeve2018deepproblog,
    author    = {Robin Manhaeve and Sebastijan Dumancic and Angelika Kimmig and Thomas Demeester and Luc De Raedt},
    title     = {{DeepProbLog: Neural Probabilistic Logic Programming}},
    booktitle = {NeurIPS},
    pages     = {3753--3763},
    year      = {2018}
}

@inproceedings{maaloe2019biva,
    author    = {Lars Maal{\o}e and Marco Fraccaro and Valentin Li{\'{e}}vin and Ole Winther},
    title     = {{BIVA: A Very Deep Hierarchy of Latent Variables for Generative Modeling}},
    booktitle = {NeurIPS},
    pages     = {6548--6558},
    year      = {2019}
}

@inproceedings{liu2021learning,
    author    = {Nan Liu and Shuang Li and Yilun Du and Joshua B. Tenenbaum and Antonio Torralba},
    title     = {{Learning to Compose Visual Relations}},
    booktitle   = {NeurIPS Workshop (CtrlGen)},
    year      = {2021}
}

@inproceedings{du2020compositional,
    author    = {Yilun Du and Shuang Li and Igor Mordatch},
    title     = {{Compositional Visual Generation and Inference with Energy Based Models}},
    booktitle   = {{NeurIPS}},
    year      = {2020}
}

@inproceedings{gregor2015draw,
    author    = {Karol Gregor and Ivo Danihelka and Alex Graves and Danilo Jimenez Rezende and Daan Wierstra},
    title     = {{DRAW:} {A} Recurrent Neural Network For Image Generation},
    booktitle = {ICML},
    volume    = {37},
    pages     = {1462--1471},
    year      = {2015}
}

@inproceedings{hong2018inferring,
    author    = {Seunghoon Hong and Dingdong Yang and Jongwook Choi and Honglak Lee},
    title     = {{Inferring Semantic Layout for Hierarchical Text-to-Image Synthesis}},
    booktitle = {CVPR},
    pages     = {7986--7994},
    year      = {2018},
}

@inproceedings{johnson2018image,
    author    = {Justin Johnson and Agrim Gupta and Li Fei{-}Fei},
    title     = {{Image Generation From Scene Graphs}},
    booktitle = {CVPR},
    pages     = {1219--1228},
    year      = {2018}
}

@inproceedings{mansimov2016generating,
    author    = {Elman Mansimov and Emilio Parisotto and Lei Jimmy Ba and Ruslan Salakhutdinov},
    title     = {Generating Images from Captions with Attention},
    booktitle = {ICLR},
    year      = {2016}
}

@inproceedings{oord2016contitional,
    author    = {A{\"{a}}ron van den Oord and Nal Kalchbrenner and Lasse Espeholt and Koray Kavukcuoglu and Oriol Vinyals and Alex Graves},
    title     = {{Conditional Image Generation with PixelCNN Decoders}},
    booktitle = {NeurIPS},
    pages     = {4790--4798},
    year      = {2016}
}

@inproceedings{ramesh2021zero,
    author    = {Aditya Ramesh and Mikhail Pavlov and Gabriel Goh and Scott Gray and Chelsea Voss and Alec Radford and Mark Chen and Ilya Sutskever},
    title     = {{Zero-Shot Text-to-Image Generation}},
    booktitle = {ICML},
    volume    = {139},
    pages     = {8821--8831},
    year      = {2021}
}

@inproceedings{reed2016generative,
    author    = {Scott E. Reed and Zeynep Akata and Xinchen Yan and Lajanugen Logeswaran and Bernt Schiele and Honglak Lee},
    title     = {{Generative Adversarial Text to Image Synthesis}},
    booktitle = {ICML},
    pages     = {1060--1069},
    year      = {2016}
}

@inproceedings{reed2016learning,
    author    = {Scott E. Reed and Zeynep Akata and Santosh Mohan and Samuel Tenka and Bernt Schiele and Honglak Lee},
    title     = {{Learning What and Where to Draw}},
    booktitle = {NeurIPS},
    year      = {2016}
}

@inproceedings{zhang2017stackgan,
    author    = {Han Zhang and Tao Xu and Hongsheng Li},
    title     = {{StackGAN: Text to Photo-Realistic Image Synthesis with Stacked Generative Adversarial Networks}},
    booktitle = {ICCV},
    pages     = {5908--5916},
    year      = {2017}
}

@article{zhang2019stackpp,
    author    = {Han Zhang and Tao Xu and Hongsheng Li and Shaoting Zhang and Xiaogang Wang and Xiaolei Huang and Dimitris N. Metaxas},
    title     = {{StackGAN++: Realistic Image Synthesis with Stacked Generative Adversarial Networks}},
    journal   = {{IEEE} Trans. Pattern Anal. Mach. Intell.},
    volume    = {41},
    number    = {8},
    pages     = {1947--1962},
    year      = {2019}
}

@article{chang2021survey,
    author={Xiaojun Chang and Pengzhen Ren and Pengfei Xu and Zhihui Li and Xiaojiang Chen and Alexander G. Hauptmann},
    journal={{IEEE} Trans. Pattern Anal. and Mach. Intell.}, 
    title={{A Comprehensive Survey of Scene Graphs: Generation and Application}},
    year={2021},
    pages={1-1},
}

@inproceedings{deng2021generative,
    author    = {Fei Deng and Zhuo Zhi and Donghun Lee and Sungjin Ahn},
    title     = {{Generative Scene Graph Networks}},
    booktitle = {ICLR},
    year      = {2021}
}

@inproceedings{oron2019scene,
    author    = {Oron Ashual and Lior Wolf},
    title     = {{Specifying Object Attributes and Relations in Interactive Scene Generation}},
    booktitle = {ICCV},
    pages     = {4560--4568},
    year      = {2019}
}

@inproceedings{jiuxiang2019scene,
    author    = {Jiuxiang Gu and Handong Zhao and Zhe Lin and Sheng Li and Jianfei Cai and Mingyang Ling},
    title     = {{Scene Graph Generation With External Knowledge and Image Reconstruction}},
    booktitle = {CVPR},
    pages     = {1969--1978},
    year      = {2019}
}

@inproceedings{herzig2020learning,
    author    = {Roei Herzig and Amir Bar and Huijuan Xu and Gal Chechik and Trevor Darrell and Amir Globerson},
    title     = {{Learning Canonical Representations for Scene Graph to Image Generation}},
    booktitle = {ECCV},
    volume    = {12371},
    pages     = {210--227},
    year      = {2020}
}

@inproceedings{hua2021relation,
    author    = {Tianyu Hua and Hongdong Zheng and Yalong Bai and Wei Zhang and Xiao{-}Ping Zhang and Tao Mei},
    title     = {{Exploiting Relationship for Complex-scene Image Generation}},
    booktitle = {AAAI},
    pages     = {1584--1592},
    year      = {2021}
}

@inproceedings{li2019paste,
    author    = {Yikang Li and Tao Ma and Yeqi Bai and Nan Duan and Sining Wei and Xiaogang Wang},
    title     = {{PasteGAN: A Semi-Parametric Method to Generate Image from Scene
               Graph}},
    booktitle = {NeurIPS},
    pages     = {3950--3960},
    year      = {2019}
}

@inproceedings{mittal2019interactive,
    author    = {Gaurav Mittal and Shubham Agrawal and Anuva Agarwal and Sushant Mehta and Tanya Marwah},
    title     = {{Interactive Image Generation Using Scene Graphs}},
    booktitle = {ICLR Workshop (DeepGenStruct)},
    year      = {2019}
}

@inproceedings{greff2017neural,
    author    = {Klaus Greff and Sjoerd van Steenkiste and J{\"{u}}rgen Schmidhuber},
    title     = {{Neural Expectation Maximization}},
    booktitle = {NeurIPS},
    year      = {2017}
}

@inproceedings{sjoerd2018relational,
    author    = {Sjoerd van Steenkiste and Michael Chang and Klaus Greff and J{\"{u}}rgen Schmidhuber},
    title     = {{Relational Neural Expectation Maximization: Unsupervised Discovery of Objects and their Interactions}},
    booktitle = {ICLR},
    year      = {2018}
}

@article{burgess2019monet,
    author    = {Christopher P. Burgess and Lo{\"{\i}}c Matthey and Nicholas Watters and Rishabh Kabra and Irina Higgins and Matthew Botvinick and Alexander Lerchner},
    title     = {{MONet: Unsupervised Scene Decomposition and Representation}},
    journal   = {CoRR},
    year      = {2019}
}

@inproceedings{engelcke2020genesis,
    author    = {Martin Engelcke and Adam R. Kosiorek and Oiwi Parker Jones and Ingmar Posner},
    title     = {{GENESIS: Generative Scene Inference and Sampling with Object-Centric Latent Representations}},
    booktitle = {ICLR},
    year      = {2020}
}

@inproceedings{locatello2020object,
    author    = {Francesco Locatello and Dirk Weissenborn and Thomas Unterthiner and Aravindh Mahendran and Georg Heigold and Jakob Uszkoreit and Alexey Dosovitskiy and Thomas Kipf},
    title     = {{Object-Centric Learning with Slot Attention}},
    booktitle = {NeurIPS},
    year      = {2020}
}

@article{kipf2021conditional,
    author    = {Thomas Kipf and Gamaleldin F. Elsayed and Aravindh Mahendran and Austin Stone and Sara Sabour and Georg Heigold and Rico Jonschkowski and Alexey Dosovitskiy and Klaus Greff},
    title     = {{Conditional Object-Centric Learning from Video}},
    journal   = {CoRR},
    year      = {2021}
}

@inproceedings{eslami2016attend,
    author    = {S. M. Ali Eslami and Nicolas Heess and Theophane Weber and Yuval Tassa and David Szepesvari and Koray Kavukcuoglu and Geoffrey E. Hinton},
    title     = {{Attend, Infer, Repeat: Fast Scene Understanding with Generative Models}},
    booktitle = {NeurIPS},
    pages     = {3225--3233},
    year      = {2016}
}

@inproceedings{crawford2019spatially,
    author    = {Eric Crawford and Joelle Pineau},
    title     = {{Spatially Invariant Unsupervised Object Detection with Convolutional Neural Networks}},
    booktitle = {AAAI},
    pages     = {3412--3420},
    year      = {2019}
}

@inproceedings{lin2020space,
    author    = {Zhixuan Lin and Yi{-}Fu Wu and Skand Vishwanath Peri and Weihao Sun and Gautam Singh and Fei Deng and Jindong Jiang and Sungjin Ahn},
    title     = {{SPACE: Unsupervised Object-Oriented Scene Representation via Spatial Attention and Decomposition}},
    booktitle = {ICLR},
    year      = {2020}
}

@inproceedings{jiang2020scalor,
    author    = {Jindong Jiang and Sepehr Janghorbani and Gerard de Melo and Sungjin Ahn},
    title     = {{SCALOR: Generative World Models with Scalable Object Representations}},
    booktitle = {ICLR},
    year      = {2020}
}

@inproceedings{greff2019iterative,
    author    = {Klaus Greff and Rapha{\"{e}}l Lopez Kaufman and Rishabh Kabra and Nick Watters and Christopher Burgess and Daniel Zoran and Loic Matthey and Matthew Botvinick and Alexander Lerchner},
    title     = {{Multi-Object Representation Learning with Iterative Variational Inference}},
    booktitle = {ICML},
    pages     = {2424--2433},
    year      = {2019}
}

@inproceedings{tulsiani2017shape,
    author    = {Shubham Tulsiani and Hao Su and Leonidas J. Guibas and Alexei A. Efros and Jitendra Malik},
    title     = {{Learning Shape Abstractions by Assembling Volumetric Primitives}},
    booktitle = {CVPR},
    pages     = {1466--1474},
    year      = {2017}
}

@article{li2017grass,
    author    = {Jun Li and Kai Xu and Siddhartha Chaudhuri and Ersin Yumer and Hao (Richard) Zhang and Leonidas J. Guibas},
    title     = {{GRASS: generative recursive autoencoders for shape structures}},
    journal   = {{ACM} Trans. Graph.},
    volume    = {36},
    number    = {4},
    pages     = {52:1--52:14},
    year      = {2017}
}

@article{zhu2018scores,
    author    = {Chenyang Zhu and Kai Xu and Siddhartha Chaudhuri and Renjiao Yi and Hao Zhang},
    title     = {{SCORES: shape composition with recursive substructure priors}},
    journal   = {{ACM} Trans. Graph.},
    volume    = {37},
    number    = {6},
    pages     = {211:1--211:14},
    year      = {2018}
}

@inproceedings{huang2020part,
    author    = {Jialei Huang and Guanqi Zhan and Qingnan Fan and Kaichun Mo and Lin Shao and Baoquan Chen and Leonidas J. Guibas and Hao Dong},
    title     = {{Generative 3D Part Assembly via Dynamic Graph Learning}},
    booktitle   = {NeurIPS},
    year      = {2020},
}

@inproceedings{kania2020ucsg,
    author    = {Kacper Kania and Maciej Zieba and Tomasz Kajdanowicz},
    title     = {{UCSG-NET - Unsupervised Discovering of Constructive Solid Geometry Tree}},
    booktitle = {NeurIPS},
    year      = {2020},
}

@inproceedings{deng2020convex,
    author    = {Boyang Deng and Kyle Genova and Soroosh Yazdani and Sofien Bouaziz and Geoffrey E. Hinton and Andrea Tagliasacchi},
    title     = {{CvxNet: Learnable Convex Decomposition}},
    booktitle = {CVPR},
    pages     = {31--41},
    year      = {2020}
}

@inproceedings{xu2019discovery,
    author    = {Zhenjia Xu and Zhijian Liu and Chen Sun and Kevin Murphy and William T. Freeman and Joshua B. Tenenbaum and Jiajun Wu},
    title     = {{Unsupervised Discovery of Parts, Structure, and Dynamics}},
    booktitle = {ICLR},
    year      = {2019}
}

@inproceedings{li2020causal,
    author    = {Yunzhu Li and Antonio Torralba and Anima Anandkumar and Dieter Fox and Animesh Garg},
    title     = {{Causal Discovery in Physical Systems from Videos}},
    booktitle = {NeurIPS},
    year      = {2020}
}

@inproceedings{stanic2021hierarchical,
    author    = {Aleksandar Stanic and Sjoerd van Steenkiste and J{\"{u}}rgen Schmidhuber},
    title     = {{Hierarchical Relational Inference}},
    booktitle = {AAAI},
    pages     = {9730--9738},
    year      = {2021}
}

@inproceedings{sabour2021capsules,
    author    = {Sara Sabour and Andrea Tagliasacchi and Soroosh Yazdani and Geoffrey E. Hinton and David J. Fleet},
    title     = {{Unsupervised Part Representation by Flow Capsules}},
    booktitle = {ICML},
    volume    = {139},
    pages     = {9213--9223},
    year      = {2021}
}

@inproceedings{gothoskar2021program,
    author    = {Nishad Gothoskar and Marco Cusumano{-}Towner and Ben Zinberg and Matin Ghavamizadeh and Falk Pollok and Austin Garrett and Joshua B. Tenenbaum and Dan Gutfreund and Vikash K. Mansinghka},
    title     = {{3DP3: 3D Scene Perception via Probabilistic Programming}},
    booktitle = {NeurIPS},
    year      = {2021},
}

@inproceedings{gopa2021keypoint,
    author    = {Anand Gopalakrishnan and Sjoerd van Steenkiste and J{\"{u}}rgen Schmidhuber},
    title     = {{Unsupervised Object Keypoint Learning using Local Spatial Predictability}},
    booktitle = {ICLR},
    year      = {2021}
}

@inproceedings{deraedt2007problog,
  title={ProbLog: A Probabilistic Prolog and Its Application in Link Discovery.},
  author={De Raedt, Luc and Kimmig, Angelika and Toivonen, Hannu},
  booktitle={IJCAI},
  volume={7},
  pages={2462--2467},
  year={2007},
}

@inproceedings{liello2020efficient,
  author    = {Luca Di Liello and
               Pierfrancesco Ardino and
               Jacopo Gobbi and
               Paolo Morettin and
               Stefano Teso and
               Andrea Passerini},
  title     = {{Efficient Generation of Structured Objects with Constrained Adversarial Networks}},
  booktitle = {NeurIPS},
  year      = {2020},
}

@inproceedings{winters2021deepstochlog,
  author    = {Thomas Winters and
               Giuseppe Marra and
               Robin Manhaeve and
               Luc De Raedt},
  title     = {DeepStochLog: Neural Stochastic Logic Programming},
  booktitle   = {UAI, to appear},
  year      = {2021}
}

@misc{kautz2020third,
  author = {Henry Kautz},
  title = {The Third AI Summer},
  howpublished = "\url{https://roc-hci.com/announcements/the-third-ai-summer/}",
  year = {2020}
}

@article{besold2017neural,
  title={Neural-symbolic learning and reasoning: A survey and interpretation},
  author={Besold, Tarek R and Garcez, Artur d'Avila and Bader, Sebastian and Bowman, Howard and Domingos, Pedro and Hitzler, Pascal and K{\"u}hnberger, Kai-Uwe and Lamb, Luis C and Lowd, Daniel and Lima, Priscila Machado Vieira and others},
  journal={arXiv preprint arXiv:1711.03902},
  year={2017}
}

@misc{bengio2021aidebate,
  author = {Bengio, Yoshua and Marcus, Gary},
  title = {AI Debate},
  howpublished = "\url{https://montrealartificialintelligence.com/aidebate/}",
  year = {2020}
}

@inproceedings{deraedt2020statistical,
  title={From statistical relational to neuro-symbolic artificial intelligence},
  author={De Raedt, Luc and Duman{\v{c}}i{\'c}, Sebastijan and Manhaeve, Robin and Marra, Giuseppe},
  booktitle={IJCAI},
  year={2020}
}

@book{kahneman2011thinking,
  title={Thinking, fast and slow},
  author={Kahneman, Daniel},
  year={2011},
  publisher={Macmillan}
}

@inproceedings{yi2018neural,
  title={Neural-Symbolic VQA: Disentangling Reasoning from Vision and Language Understanding},
  author={Yi, Kexin and Wu, Jiajun and Gan, Chuang and Torralba, Antonio and Kohli, Pushmeet and Tenenbaum, Joshua B},
  booktitle = {NeurIPS},
  year = {2018}
}

@inproceedings{minervini2020learning, title={Learning reasoning strategies in end-to-end differentiable proving},
author={Minervini, Pasquale and Riedel, Sebastian and Stenetorp, Pontus and Grefenstette, Edward and Rockt{\"a}schel, Tim},
booktitle={ICML},
year={2020},
}
